\documentclass[11pt]{article}

\usepackage{mathrsfs}
\usepackage{longtable}
\usepackage{makeidx}
\usepackage{fullpage}
\usepackage{amsmath}
\usepackage{amssymb}
\usepackage{setspace}
\usepackage{bbm}
\usepackage{dsfont}
\usepackage{graphics}
\usepackage{epsfig}
\usepackage[font=footnotesize,labelfont=bf,justification=centerlast,width=.94\textwidth]{caption}
 \usepackage{multirow}
\usepackage{array,booktabs}
\usepackage{cite}
\usepackage{color}

\usepackage{hyperref}

\hypersetup{
    bookmarks=true,%
    colorlinks,%
    citecolor=blue,%
    filecolor=black,%
    linkcolor=black,%
    urlcolor=blue
}

\newcommand{\be}{\begin{equation}}
\newcommand{\ee}{\end{equation}}
\newcommand{\bes}{\begin{equation*}}
\newcommand{\ees}{\end{equation*}}
\newcommand{\bea}{\begin{eqnarray}}
\newcommand{\eea}{\end{eqnarray}}
\newcommand{\beas}{\begin{eqnarray*}}
\newcommand{\eeas}{\end{eqnarray*}}

\newcommand{\bmat}{\begin{bmatrix}}
\newcommand{\emat}{\end{bmatrix}}

\def\le{\left}
\def\ri{\right}

\def\le{\left}
\def\ri{\right}

\def\CT{{\cal T}}

\begin{document}
\numberwithin{equation}{section}
{
\begin{titlepage}
\begin{center}

\hfill \\
\hfill \\
\vskip 0.75in

{\Large \bf Geodesic Diagrams, Gravitational Interactions \& OPE  Structures}\\

\vskip 0.4in

{\large Alejandra Castro, Eva Llabr\'es, and Fernando Rejon-Barrera}\\

\vskip 0.3in

{\it Institute for Theoretical Physics Amsterdam and Delta Institute for Theoretical Physics, University of Amsterdam, Science Park 904, 1098 XH Amsterdam, The Netherlands} \vskip .5mm

\texttt{a.castro@uva.nl, e.m.llabres@uva.nl, f.g.rejonbarrera@uva.nl}

\end{center}

\vskip 0.35in

\begin{center} {\bf ABSTRACT } \end{center}
We give a systematic procedure to evaluate conformal partial waves involving symmetric tensors for an arbitrary CFT$_d$ using geodesic Witten diagrams in AdS$_{d+1}$. Using this procedure we discuss how to draw a line between the tensor structures in the CFT  and cubic interactions in AdS.  We contrast this map to known results using three-point Witten diagrams: the maps obtained via volume versus geodesic integrals differ.  Despite these differences, we  show how  to decompose four-point exchange Witten diagrams in terms of geodesic diagrams, and we discuss the product expansion of local bulk fields in AdS. 
\vfill

\noindent \today

\end{titlepage}
}
%%%%%%%%%%%%%%%%%%%%%%%%%%%%%%%%%%%%%%%%%%%%%%%%%%%%%%%%%%%%%%%%%%

%%%%%%%%%%%%%%%%%%%%%
\newpage

\tableofcontents

\section{Introduction}

Conformal field theories (CFTs) have a unique position within quantum field theory. They are central to the ambitious questions that drives many theorists: the quest of classifying  all possible  fixed points of the renormalization group equations, and unveiling the theorems that accompany the classification. And in modern times, they are also at the center of the holographic principle. Conformal field theories are  key to unveil novel features about quantum gravity in AdS.  

In a CFT symmetries play a crucial role. The exploitation of the conformal group gives an efficient organizational principle for the observables in the theory. In particular, the conformal block decomposition of four point correlation functions is such a principle: it is natural to cast the four point function into portions that are purely determined by symmetries (conformal partial waves) and the theory dependent data (OPE coefficients). Having analytic and numerical control over this decomposition has been key in  recent  developments. This includes the impressive revival of the conformal bootstrap program \cite{Rattazzi2008,Poland2012,El-Showk2012}, and we refer to  \cite{Rychkov2016,Simmons-Duffin2016,Poland2016}  for an overview on this  area.  

Our aim here is to apply the efficiency of the conformal block decomposition to holography: can we organize observables in AdS gravity as we do in a CFT? This question has been at the heart of holography since its conception \cite{Maldacena1999,Gubser1998,Witten1998},  with perhaps the most influential result the prescription to evaluate CFT correlation functions via Witten diagrams \cite{Witten1998}. But only until very recently the concept of conformal partial wave was addressed directly in holography: the authors in  \cite{Hijano2016} proposed that the counterpart of a CFT$_d$ conformal partial wave is a \emph{geodesic Witten diagram} in AdS$_{d+1}$. As for the conventional Witten diagram it involves bulk--to--boundary and bulk--to--bulk propagators in AdS, with the important difference that the contact terms of the fields are projected over geodesics rather than integrated over the entire volume of AdS. Among the many results presented in \cite{Hijano2016} to support their proposal, they reproduced explicitly the scalar conformal partial waves in a CFT$_d$\footnote{We use the term  \emph{scalar} conformal partial wave to denote that the external fields are scalar operators; the exchanged field can be a symmetric traceless tensor.  A \emph{spinning} conformal partial wave is when at least one external fields is a symmetric traceless tensor. } via a geodesic diagram in AdS. %Along similar lines, there is also a proposal to describe Virasoro blocks via geodesic diagrams \cite{Hijano2015}, but here we will only discuss global conformal blocks.

The goal of this paper is twofold: to give a method to evaluate a spinning conformal partial wave using holography, and to show how Witten diagrams decompose in terms of these building blocks. The first step towards this direction was given \cite{Nishida2016}, where only one external leg had non-trivial spin. Here we expand that discussion to include spin on all possible positions of the diagram, and our current limitation is that we are only considering symmetric and traceless fields in the external and exchange positions. Our strategy is to cast the CFT construction of conformal partial waves in \cite{Costa2011} along the lines of the AdS proposal in \cite{Hijano2016}.  In particular, we will show how to decode the tensor structures (i.e. OPE structures) appearing in three point functions and conformal partial waves in terms of bulk differential operators acting on geodesic diagrams.

 Witten diagrams,  that are in any way more complicated than those with three legs and tree level, are infamous for how difficult it is to evaluate them. The integrals involved become quite cumbersome as the specie of the field changes, and even more intricate if internal lines are involved. The first explicit results are those in \cite{LiuTseytlin1999,Liu1999,FreedmanMathurMatusisEtAl1999,DHokerMathurMatusisEtAl2000,DHokerFreedman1999,DHokerFreedmanMathurEtAl1999,HoffmannPetkouRuhl2002,HoffmannPetkouRuhl2000}, and more recently the subject has been address by using a Mellin decomposition of the diagrams (see e.g.  \cite{Penedones2011,Paulos2011,Fitzpatrick2011,Costa2012,Fitzpatrick2012,Goncalves2015,Gopakumar2016}). Having a clean and efficient decomposition of a Witten diagram in terms of geodesic diagrams is a computational tool that can allow  a new level of precision in holography. Our method to decode the tensor structures provides a novel step forward in this direction by optimizing the evaluation of correlations functions in AdS/CFT.

A good portion of our analysis will involve the map between tensor structures in the CFT and cubic interactions in AdS. And in this arena there are already universal results in higher spin holography.  One of the goals in that field is to understand locality and effective Lagrangians within Vasiliev's higher spin theory. And in this context various quartic and cubic interactions have been successfully mapped to their counterpart in the CFT \cite{Petkou2003,Erdmenger2014,Bekaert2015,Bekaert2015a,Skvortsov2017,Sleight2016b,Sleight2016a}. The extent of their results is by no means limited to higher spin gravity. One impressive part of the literature is the identification of all independent structures of cubic vertices for either massive, massless or partially massive cubic interactions of symmetric traceless tensors \cite{Metsaev2006,Joung2012,Taronna2012,Joung2013,Joung2013a}. The other impressive side of this literature is the precise identification of each cubic interaction with a tensor structure of the CFT \cite{Sleight2016b}. We will use the results there in two ways.  First, we will contrast the tensor structures that a geodesic diagram captures versus the analog  Witten diagram:  this puts these diagrams in a very different footing when it comes to capturing dynamical properties of AdS rather than objects designed to be driven  purely  by symmetries.  Second, we will use the identities developed in \cite{Sleight2016b} to argue that a four-point exchange Witten diagram can be decomposed in terms of geodesic integrals.

This paper is organized as follows. Section \ref{sec:esp} is a review on the embedding space formalism to describe CFT$_d$ and AdS$_{d+1}$ quantities. In particular in section \ref{sec:cftcorr} we will review the classification of OPE structures in  CFT, and how they are obtained via suitable differential operators. Our main result is in section \ref{sec:maingwd} where we give an AdS counterpart of the operators in section \ref{sec:cftcorr}. This shows how one can obtain any spinning conformal partial wave via an appropriate geodesic Witten diagram with perfect agreement with the CFT. In section \ref{sec:interactions} we discuss certain features of this method by focusing mostly on low spin examples. We first discuss the relation among gravitational interactions and OPE structures using geodesic diagrams, and contrast it with the reconstruction done using   Witten diagrams. Even though there are non-trivial cancellations in the geodesic diagrams (which do not occur with volume integrals), in section \ref{sec:decom} we show how to decompose four point exchange Witten diagrams in terms of geodesic diagrams. We end with a discussion of our results and future directions in section \ref{sec:disc}.

{\bf Note Added:} At the same time this work was completed, in \cite{Dyer:2017zef,Chen:2017yia} the authors also address the question of how to capture spinning conformal partial waves in terms of geodesic Witten diagrams. 
  
\section{Embedding space formalism}   \label{sec:esp}

The simplest way to carry out our analysis is via the embedding space formalism. We will use this to describe both CFT$_d$ and AdS$_{d+1}$ quantities. This formalism was recently revisited and exploited  in \cite{Costa2011b,Costa2014,Sleight2016b}, and we mainly follow their presentation.  This section summarises the most important definitions and relations we will use throughout; readers familiar with this material can skip this section.  All of our discussion will be in Euclidean signature.

 \subsection{CFT side of embedding}\label{sec:cft}
A natural description of the conformal group $SO(d+1,1)$ is  in the embedding space $\mathbb{M}^{d+2}$: this makes conformal symmetry constraints simple Lorentz symmetry conditions (which are more easily implemented). In this section we will show  how to uplift the CFT$_d$ fields on $\mathbb{R}^{d}$ to $\mathbb{M}^{d+2}$, and write correlation functions in this language.

 The dot product in $\mathbb{M}^{d+2}$ is given by 
 \begin{align}
 P_1\cdot P_2\equiv P_1^{A}P_2^{B}\eta_{AB}=-\frac12 P_1^{+}P_2^{-}-\frac12 P_1^{-}P_2^{+}+\delta_{ab}P_1^{a}P_2^{b}~,
 \end{align}
 where we are using light cone coordinates 
 \begin{align}
 P^{A}=(P^{+},P^{-},P^{a})~.
 \end{align}
A point $x^{a}\in \mathbb{R}^{d}$ is embedded in $\mathbb{M}^{d+2}$ by null stereographic map of the  coordinates 
 \begin{align}\label{eq:px}
 x^{a}~\to~ P^{A}=(1,x^{2},x^{a})~, \qquad a=1,\ldots, d~.
 \end{align}
 This implies that the CFT$_d$ coordinates live in the projective light cone
 \be\label{eq:light}
 P^{2}=0 ~, \quad P\equiv \lambda P~, \quad \lambda \in \mathbb{R}~.
 \ee

In the embedding formalism there is a very economical way of manipulating  $\mathbb{R}^{d}$ symmetric and traceless tensors.   This is discussed extensively in \cite{Costa2011b}, and the bottom line is to encode the tensorial properties in a polynomial.  One defines an auxiliary vector $Z^A$, and considers the contraction
 \begin{align}\label{eq:cfttensor}
 T(P,Z)\equiv Z^{A_{1}}\cdots Z^{A_{n}}T_{A_{1}\cdots A_{n}}(P)~,
 \end{align}
 with the following restrictions and properties:
 \begin{enumerate}
 \item $Z^2=0$ encodes the traceless condition.
 \item $T(P,Z+\alpha P)=T(P,Z)$  makes the tensor tangent to the light cone $P^2=0$. 
 \item Homogeneity defines the conformal weight $\Delta$ and spin $l$ as $T(\lambda P, \alpha Z)= \lambda^{-\Delta} \alpha^l T( P, Z)$. 
 \end{enumerate}
All of these condition are conformally invariant which makes $T_{A_{1}\cdots A_{n}}(P)$ an $SO(d+1,1)$ symmetric traceless tensor. From here, a symmetric traceless tensor field on $\mathbb{R}^d$ is given by
\be\label{eq:projR}
t_{a_{1}\cdots a_{n}}= {\partial P^{A_1}\over \partial x^{a_1}}\ldots  {\partial P^{A_n}\over \partial x^{a_n}}T_{A_{1}\cdots A_{n}}(P)~,
\ee
with $P^A$ given by \eqref{eq:px}. It is important to note that  any tensor $T_{A_{1}\cdots A_{n}}(P)$ proportional to $P^A$ projects to zero: such tensor will be \emph{pure gauge}. And  hence, without loss of generality we can require the orthogonality condition
\be
Z\cdot P=0~.
\ee

We can as well extract $t_{a_{1}\cdots a_{n}}$ from the polynomial directly. First, the polynomial in $(d+2)$-dimensions can be brought into $d$-dimensional variables via the relation
\be
T(P,Z) = t(x, z) ~, \quad {\rm with}  \quad Z^A=(0,2 x\cdot z,z) ~, \quad  P^{A}=(1,x^{2},x^{a})~.
\ee
Then the components of the tensor in $\mathbb{R}^d$ are
\be\label{eq:projD}
t_{a_{1}\cdots a_{n}} = {1\over n! (d/2-1)_n }D_{a_1}\ldots D_{a_n} t(x, z)~,
\ee
where $(d)_l=\Gamma(d+l)/\Gamma(d)$ and $D_a$ are differential operators that do the job of projecting the polynomial to symmetric traceless tensors:
\be
D_a=\le({d\over 2}-1 +z\cdot {\partial \over \partial z}\ri){\partial \over \partial z^a} -{1\over 2} z^a {\partial^2 \over \partial z\cdot \partial z}~.
\ee
This operator is also convenient for other purposes. For example, we can do full contractions via the polynomial directly: given two symmetric traceless tensors in $\mathbb{R}^d$, their contraction is 
\be
f_{a_{1}\cdots a_{n}} g^{{a_{1}\cdots a_{n}}} = {1\over n! (d/2-1)_n } f(x,D) g(x,z)~.
\ee
In the $(d+2)$-dimensional variables we have
\be
f_{a_{1}\cdots a_{n}} g^{{a_{1}\cdots a_{n}}} ={1\over n! (d/2-1)_n } F(P,D) G(P,Z)~,
\ee
where 
\be\label{eq:da1}
D_A=\le({d\over 2}-1 +Z\cdot {\partial \over \partial Z}\ri){\partial \over \partial Z^A} -{1\over 2} Z_A {\partial^2 \over \partial Z\cdot \partial Z}~.
\ee

\subsubsection{CFT$_d$ correlation functions}\label{sec:cftcorr}
The main appeal of the embedding formalism is that one can conveniently describe $n$-point functions for symmetric tensors which automatically satisfy the constraints of $SO(d+1,1)$. In a nutshell, the task ahead is to identify polynomials in $(P_i,Z_j)$ of the correct homogeneity modulo terms of order $Z_i^2$ and $Z_i\cdot P_i$.  

To start, consider the two point function of a spin $l$ primary of conformal dimension $\Delta$ in embedding space. This correlation function is a $2l$ tensor which we encode in a polynomial as
\be
G_{\Delta|l}(P_1,Z_1; P_2, Z_2)\equiv Z_1^{A_1}\ldots Z_1^{A_l}Z_2^{B_1}\ldots Z_2^{B_l} G_{A_1\ldots A_l B_1\ldots B_l } (P_1,P_2)~,
\ee
and projecting further to $\mathbb{R}^d$ is done via \eqref{eq:projR} or \eqref{eq:projD}. Up to a constant, the appropriate polynomial is
\be\label{eq:2pcft}
G_{\Delta|l}(P_1,Z_1; P_2, Z_2)= {(H_{12})^l\over (P_{12})^{\Delta}}~,
\ee
where we have introduced
\be
P_{12}\equiv -2P_1\cdot P_2~, \qquad H_{12}(Z_1,Z_2)\equiv Z_{1}\cdot Z_{2}+2 \frac{(Z_{1}\cdot P_{2}) (Z_{2}\cdot P_{1})}{P_{12}}~.
\ee
The numerator in \eqref{eq:2pcft} assures that we have a polynomial of degree $l$ (encoding the tensorial features), while the denominator contains the homogeneity property we expect from conformal invariance. One can check as well that all other properties listed below \eqref{eq:cfttensor} are satisfied, and the solution is unique up to pure gauge terms. 

 Three point functions  of symmetric traceless operators have an elegant description in this language as well. Consider three primaries of conformal dimension $\Delta_i$ and spin $l_i$: the three point function is expected to take the form
 \be\label{eq:3pt}
 G_{\Delta_1,\Delta_2,\Delta_3|l_1,l_2,l_3}(P_i,Z_i)= {Q_3(P_i,Z_i)\over (P_{12})^{(\Delta_1+\Delta_2 -\Delta_3)/2} (P_{23})^{(\Delta_2+\Delta_3 -\Delta_1)/2}(P_{13})^{(\Delta_1+\Delta_3 -\Delta_2)/2}  }~.
 \ee
 The denominator is chosen such that the homogeneity with respect to $P_i$ is explicit. The numerator $Q_3$ should be a transverse polynomial of degree $l_i$ for each $Z_i$, and homogenous of degree zero for each $P_i$. Given these properties, we can cast the desired polynomial in terms of 6 building blocks \cite{Costa2011b}:\footnote{Our conventions for  $V_{i,jk}$ and  $H_{ij}$ are very similar to those in \cite{CostaHansenPenedonesEtAl2016}, which differ slightly from those in \cite{Costa2011b}. Note that our definition of $V_{i,jk}$ differs to that of \cite{CostaHansenPenedonesEtAl2016} by a minus sign.}
 \begin{align}
&V_{1,23}~,\quad  V_{2,31}~, \quad V_{3,21} ~,\cr
 &H_{12}~, \quad H_{13}~, \quad H_{23}~,
 \end{align}
where
 \begin{align}\label{eq:VH}
 V_{i,jk}=& {(Z_i\cdot P_j)P_{ik}-(Z_i\cdot P_k)P_{ij}\over \sqrt{P_{ij} P_{ik} P_{jk}}}~,\cr
 H_{ij}=&\, Z_{i}\cdot Z_{j}+2 \frac{(Z_{j}\cdot P_{i}) (Z_{i}\cdot P_{j})}{P_{ij}}~.
 \end{align}
 $Q_3$ then takes the general form
 \be\label{eq:q3}
 Q_3(P_i,Z_i)= \sum_{n_i\geq 0}  C_{n_1,n_2,n_3}(V_{1,23})^{l_1-n_2-n_3}(V_{2,31})^{l_2-n_3-n_1} (V_{3,21})^{l_3-n_1-n_2}  H_{12}^{n_1} H_{13}^{n_3} H_{23}^{n_2}~,
 \ee
 giving us the expected homogeneity and transverse properties. Here $C_{n_1,n_2,n_3}$ are constant (theory dependent) coefficients. Note that each of the powers of $V_{i,jk}$ in \eqref{eq:q3} have to be positive, and this restricts the number of possible combinations. For fixed $l_i$ the number of tensorial structures is
 \be\label{eq:n123}
 N(l_1,l_2,l_3)= {1\over 6}(l_1+1)(l_1+2)(3l_2-l_1+3)-{1\over 24}p(p+2)(2p+5)-{1\over 16}(1-(-1)^p)~,
 \ee
with $l_1\leq l_2\leq l_3$ and $p\equiv {\rm max}(0,l_1+l_2-l_3)$.

For operational purposes, and later on to evaluate conformal partial waves, it is more convenient to generate the tensorial structures in \eqref{eq:q3} via differential operators. This was originally done in \cite{Costa2011}, and the basic idea is as follows. Say we look at the OPE of two operators which carry spin:
\be
{\cal O}_1^{l_1}(x_1){\cal O}_2^{l_2}(x_2)=\sum_{{\cal O}} \lambda_{12{\cal O}}C(x_{12},\partial_2)^{l_1,l_2,l_3} \, {\cal O}_3^{l_3}(x_2) ~.
\ee
The OPE structures now carry the tensorial properties of the external operators, relative to cases where the left hand side operators are scalar primaries. The point made in \cite{Costa2011} is to view these more complicated objects as derivatives of the basic scalar OPE. More explicitly, if the OPE between two scalar primaries is
\be\label{eq:ope12}
{\cal O}_1(x_1){\cal O}_2(x_2)=\sum_{{\cal O}_3} \lambda_{12{\cal O}}C(x_{12},\partial_2)^{l_3} \, {\cal O}_3^{l_3}(x_2) ~,
\ee
then
\be
C(x_{12},\partial_2)^{l_1,l_2,l_3} = D_{x_1,x_2}^{l_1,l_2} C(x_{12},\partial_2)^{l_3}~,
\ee
where $D_{x_1,x_2}^{l_1,l_2} $ is a differential operator that creates the tensorial structure for $l_1$ and $l_2$. Taking this relation for granted, it would then imply that the three point functions would be related as
\be\label{eq:odo}
\langle{\cal O}_1^{l_1}(x_1){\cal O}_2^{l_2}(x_2) {\cal O}_3^{s_3}(x_3) \rangle =D_{x_1,x_2}^{l_1,l_2}\langle{\cal O}_1(x_1){\cal O}_2(x_2) {\cal O}_3^{l_3}(x_3) \rangle~.
\ee
The idea is that we can represent any three point function of symmetric traceless structures as derivatives of a scalar-scalar-spin correlation function. 

One can cast as well \eqref{eq:odo} as a polynomial relation in embedding space: given a function $G_{\Delta_1,\Delta_2,\Delta_3|l_1,l_2,l_3}(P_i,Z_i)$ of certain degree in $Z_i$, we would like to relate it to a polynomial of lower degree via suitable differential operators, i.e.
\be
G_{\Delta_1,\Delta_2,\Delta_3|l_1,l_2,l_3} = D\le(P_{i}, Z_{i},{\partial \over \partial P_i}, {\partial \over \partial Z_i}\ri)G_{\Delta'_1,\Delta'_2,\Delta_3|0,0,l_3} + O(Z_{i}^2, P_{i}^2, Z_i\cdot P_i)~,\quad i=1,2~.
\ee
The differential operators have to satisfy certain basic properties:
\begin{enumerate}
\item $D$ must raise the degree in $Z_1$ up to $l_1$ and $Z_2$ up to $l_2$.
\item $D$ must take terms $O(Z_{n}^2, P_{n}^2, Z_n\cdot P_n)$ to terms of the same kind: keep pure gauge terms as pure gauge.
\item $D$ must map transverse functions to themselves.  
\end{enumerate}
 A basis of operators that will satisfy these requirements are
\begin{align}\label{eq:dij}
% D_{ii}=
D_{1\, ij}\equiv &- \frac12 P_{ij}\left(Z_{i}\cdot \frac{\partial}{\partial P_{j}}\right)-\left(Z_{i}\cdot P_{j} \right)\left( P_{i}\cdot \frac{\partial}{\partial P_{j}} \right)
-\left(Z_{i}\cdot Z_{j} \right)\left( P_{i}\cdot \frac{\partial}{\partial Z_{j}} \right) +\left(Z_{j}\cdot P_{i} \right)\left( Z_{i}\cdot \frac{\partial}{\partial Z_{j}} \right)~, \notag\\
% D_{ij}=
D_{2\, ij}\equiv &-\frac12 P_{ij}\left(Z_{i}\cdot \frac{\partial}{\partial P_{i}}\right)-\left(Z_{i}\cdot P_{j} \right)\left( P_{i}\cdot \frac{\partial}{\partial P_{i}} \right)
 +\left(Z_{i}\cdot P_{j} \right)\left( Z_{i}\cdot \frac{\partial}{\partial Z_{i}} \right) ~,
\end{align}
in addition to $H_{ij}$ in \eqref{eq:VH}. The operator $D_{1\, ij}$ increases the spin at position $i$ by one and decreases the dimension by one at position $i$; $D_{2\, ij}$ increases the spin at position $i$ by one and decreases the dimension by one at position $j$. $H_{ij}$ increases the spin by one at both $i$ and $j$ and leaves the conformal dimensions unchanged. 
 The commutation relation between these operators are
\begin{align}
[D_{1\, 12},D_{1\, 21}]&=\frac12 P_{12}H_{12}\left( Z_{1}\cdot \partial_{Z_{1}}-Z_{2}\cdot \partial_{Z_{2}} +P_{1}\cdot \partial_{P_{1}}-P_{2}\cdot \partial_{P_{2}}\right)~ ,\\
[D_{2\, 12},D_{2\, 21}]&=\frac12 P_{12}H_{12}\left( Z_{1}\cdot \partial_{Z_{1}}-Z_{2}\cdot \partial_{Z_{2}} -P_{1}\cdot \partial_{P_{1}}+P_{2}\cdot \partial_{P_{2}}\right)~,
\end{align}
and all other pairings are zero, including $[D_{k\,ij},H_{i'j'}]=0$.

To see how this works, it is useful to just state the map for a few examples. Defining the three point function of three scalar primaries as
\be\label{eq:3scalars}
T(\Delta_1,\Delta_2,\Delta_3)\equiv  {1\over (P_{12})^{(\Delta_1+\Delta_2 -\Delta_3)/2} (P_{23})^{(\Delta_2+\Delta_3 -\Delta_1)/2}(P_{13})^{(\Delta_1+\Delta_3 -\Delta_2)/2}  }
\ee
we have that increasing the spin by one at position $i=1$ is achieved by
\begin{align}\label{eq:vss}
G_{\Delta_1,\Delta_2,\Delta_3|1,0,0}&= V_{1,23} T(\Delta_1,\Delta_2,\Delta_3)\cr
&= {2\over \Delta_3+\Delta_2-\Delta_1-1} D_{1\,12} T(\Delta_1+1,\Delta_2,\Delta_3) \cr
&= {2\over \Delta_3-\Delta_2+\Delta_1-1} D_{2\,12} T(\Delta_1,\Delta_2+1,\Delta_3)~.
\end{align}
In the first line we wrote it as in \eqref{eq:3pt}-\eqref{eq:q3}, and in the last two lines we casted the same answer in terms of differential operators acting on the scalar correlation function. The three point function of two vectors and a scalar is  the superposition of two tensorial structures:
\be
G_{\Delta_1,\Delta_2,\Delta_3|1,1,0}= C_{1} V_{1,23}V_{2,13} T(\Delta_1,\Delta_2,\Delta_3) +C_2 H_{12}T(\Delta_1,\Delta_2,\Delta_3)~.
\ee
The first term can be written in terms of derivatives as
\begin{align} \label{eq:vv}
V_{1,23}V_{2,13} T(\Delta_1,\Delta_2,\Delta_3)&={4\over \Delta_3^2- (\Delta_1-\Delta_2)^2} D_{1\,12} D_{1\, 21}T(\Delta_1+1,\Delta_2+1,\Delta_3) \cr &+{H_{12}\over \Delta_3+\Delta_2-\Delta_1} T(\Delta_1,\Delta_2,\Delta_3)~.
\end{align}
How to map the polynomials $V_{i,jk}$'s to $D_{i\,jk}$'s is not one-to-one, as reflected explicitly in \eqref{eq:vss} among other cases. Nevertheless, one can always go from the basis of $V_{i,jk}$'s to $D_{i\,jk}$'s, and this transformation can be implemented systematically as discussed in \cite{Costa2011}. In appendix \ref{app:tensors} we give further examples and discuss briefly the conditions on $Q_3$ imposed by conservation. 

An interesting application of these differential operators is to evaluate conformal partial waves as done in \cite{Costa2011}. Given the four point function of four scalar primaries, the  conformal partial wave  decomposition is defined as \cite{Ferrara1971a,Ferrara1972,Ferrara1975}
\be\label{eq:cpw}
\langle {\cal O}_1(x_1){\cal O}_2(x_2) {\cal O}_3(x_3){\cal O}_4(x_4)\rangle = \sum_{{\cal O}} \lambda_{12{\cal O}}\lambda_{34{\cal O}}W_{\Delta | l}(x_1,x_2,x_3,x_4)~,
\ee
where $ \lambda_{ij{\cal O}}$ are theory dependent constant coefficients,  and  ${\cal O}$ is a primary of conformal dimension $\Delta$ and spin $l$. The sum over all operators $\cal O$ that appear in the OPE of  ${\cal O}_1(x_1){\cal O}_2(x_2)$. $W_{\Delta | l}(x_1,x_2,x_3,x_4)$ is known as a conformal partial wave, which is mostly characterised by  the properties of ${\cal O}$, and otherwise determined by conformal invariance and the quantum numbers of ${\cal O}_i$. In embedding space we have
\be
W_{\Delta | l}(P_1,P_2,P_3,P_4)= \le({P_{24}\over P_{14}}\ri)^{(\Delta_{1}-\Delta_2)/2}\le({P_{14}\over P_{13}}\ri)^{(\Delta_{3}-\Delta_4)/2}{G_{\Delta | l}(u,v)\over (P_{12})^{(\Delta_1+\Delta_2)/2}(P_{34})^{(\Delta_3+\Delta_4)/2}}~,
\ee
with 
\be
 u\equiv {P_{12}P_{34}\over P_{13}P_{24}}~,\quad v\equiv {P_{14}P_{23}\over P_{13}P_{24}}~.
\ee
$G_{\Delta | l}(u,v)$ is known as a conformal block, and explicit expressions can be found in e.g. \cite{DolanOsborn2011,DolanOsborn2004,DolanOsborn2001} among many other places. If we wanted to build now a conformal partial wave when the external operators are symmetric traceless tensor, the way has been paved from the discussion around \eqref{eq:ope12}-\eqref{eq:odo}. As a result of the analysis for three point functions, the partial waves of non-zero spin $l_i$ operators is simply derivatives acting on the known scalar partial wave, i.e.
\be\label{eq:cpws}
W_{\Delta | l}^{l_1,l_2,l_3,l_4}(x_1,x_2,x_3,x_4)=D_{x_1,x_2}^{l_1,l_2}D_{x_3,x_4}^{l_3,l_4}W_{\Delta | l}(x_1,x_2,x_3,x_4)~.
\ee
And in the embedding space formalism, the conformal partial wave is a suitable polynomial with the basis of differential operators that generate the tensor structures are given by \eqref{eq:dij} and $H_{ij}$. More explicitly 
 \be\label{eq:cpwdiff}
W_{\Delta | l}^{l_1,l_2,l_3,l_4}(P_i;Z_i)=D_{\rm left}D_{\rm right}W_{\Delta | l}(P_1,P_2,P_3,P_4)~,
\ee
with  $D_{\rm left}$ is a chain of powers of $D_{i\, jk}$ and $H_{ij}$ operators acting on $(P_1,P_2)$, and similarly for $D_{\rm right}$  acting on $(P_3,P_4)$. The exchange field $\cal O$  is neccesarly a traceless symmetric tensor.

%%%%%%%%%%%%%%%%%%%%%%%%%%%%%%%%%%%%%%%%%%%%%%%%%%%%%%%%%%%%%%%%%%%%%
%%%%%%%%%%%%%%%%%%%%%%%%%%%%%%%%%%%%%%%%%%%%%%%%%%%%%%%%%%%%%%%%%%%%%
\subsection{AdS side of embedding}

 The embedding formalism is as well incredibly useful to encode tensorial structures in AdS. Here we will follow \cite{Taronna2012,Costa2014}, and we highlight \cite{Bekaert2015a,Bekaert2015,Erdmenger2014,Sleight2016b} for its recent use in the context of higher spin gravity.  Euclidean AdS$_{d+1}$ in Poincare coordinates is given by 
 \be
 ds^2_{\rm AdS}={1\over r^2}\le( {dr^2  + dx^a dx_a}\ri)~.
 \ee
For sake of simplicity we are taking the AdS radius to be one. From the perspective of $\mathbb{M}^{d+2}$, AdS$_{d+1}$ is the future directed hyperboloid, i.e.  
\be
Y^2=-1 ~,\quad Y^0>0~,\qquad Y\in \mathbb{M}^{d+2}~.
\ee
This condition mapped to Poincare coordinates reads
 \begin{align}\label{eq:defny}
y^\mu =(r,x^{a})~\to ~Y^{A}=\frac{1}{r}(1,r^{2}+x^{2},x^{a})~.
 \end{align}
The AdS boundary points are obtained by sending $Y\to \infty$, and in this limit we approach the light cone \eqref{eq:light}. The induced AdS metric is
\be
G_{AB}=\eta_{AB}+Y_A Y_B~,
\ee
which plays a role as a projector.

Following the CFT discussion, we can as well describe symmetric and traceless tensor in AdS$_{d+1}$  as polynomials \cite{Costa2014}. Adapting the conditions in \eqref{eq:cfttensor} to AdS gives
 \begin{align}\label{eq:polenc}
 \CT(Y;W)\equiv W^{A_{1}}\cdots W^{A_{n}}\CT_{A_{1}\cdots A_{n}}(Y)~,
 \end{align}
where we introduce now a auxiliary tensor $W^A$. The restrictions and properties are
 \begin{enumerate}
 \item $W^2=0$ encodes the traceless condition.
 \item $W\cdot Y=0$  imposes an orthogonality condition. 
 \item Requiring that $\CT(Y, W+\alpha Y)=\CT(Y,W)$ makes the tensor transverse to the surface $Y^2=-1$. 
 \item Homogeneity $(Y\cdot \partial_Y+ W\cdot \partial_W +\mu)\CT(Y,W)=0 $ for some given value of $\mu$.\footnote{For a bulk massive spin-$J$ field in AdS$_{d+1}$, we have $\mu=\Delta+J$ with $M^2=\Delta(\Delta-d)-J$.}
 \end{enumerate}
The components of the tensor can be easily recovered by introducing a projector. Given
\begin{align}
K_A&={d-1\over 2} \le({\partial\over \partial W^A}+Y_A Y\cdot{\partial\over \partial W}\ri)+ W\cdot {\partial\over \partial W} \,{\partial\over \partial W^A}  \cr 
&\quad+ Y_A \le(W\cdot {\partial\over \partial W}\ri) \le(Y\cdot {\partial\over \partial W}\ri) -\frac12 W_{A}\le( {\partial^2\over \partial W\cdot\partial W}+ Y\cdot {\partial\over \partial W} \,Y\cdot {\partial\over \partial W}\ri)~,
\end{align}
%
%\begin{align}
%K_A=\le({d-1\over 2}+W\cdot {\partial\over \partial W}\ri){\partial\over \partial W^A}~,
%\end{align}
%
we obtain symmetric and traceless tensor in AdS via
\be
\CT_{A_{1}\cdots A_{n}}(Y)={1\over n! \le({d-1\over 2}\ri)_n}K_{A_{1}}\cdots K_{A_{n}}\CT(Y,W)~.
\ee
And the component in AdS$_{d+1}$ space is
\be\label{eq:projads}
t_{\mu_{1}\cdots \mu_{n}}= {\partial Y^{A_1}\over \partial y^{\mu_1}}\ldots  {\partial Y^{A_n}\over \partial y^{\mu_n}}{\cal T}_{A_{1}\cdots A_{n}}(Y)~,
\ee
If a tensor is of the type ${\cal T}_{A_{1}\cdots A_{n}}(Y)=Y_{(A_1}{\cal T}_{A_{2}\cdots A_{n})}(Y)~$  it is unphysical, i.e. it has a vanishing projection to AdS$_{d+1}$.

A covariant derivative in AdS is defined in the ambient space  $\mathbb{M}^{d+2}$ as
\be
\nabla_A = {\partial\over \partial Y^A} + Y_A \le(Y\cdot {\partial\over \partial Y}\ri) + W_A \le(Y\cdot {\partial\over \partial W}\ri)~.
\ee
When acting on an transverse tensor we have
\be
\nabla_B {\cal T}_{A_{1}\cdots A_{n}}(Y)= G_B^{B_1}G_{A_1}^{C_1}\cdots G_{A_n}^{C_n} {\partial\over \partial Y^{B_1}} {\cal T}_{C_{1}\cdots C_{n}}(Y)~,
\ee
where $G_{AB}$ is the induced AdS metric. Using the polynomial notation, we can write the divergence of a tensor as
\be
\nabla\cdot (K \CT(Y,W)) ~,
\ee
which after projecting to AdS$_{d+1}$ would give $\nabla^\mu {t}_{\mu \mu_2\ldots \mu_n}$. And we can as well write
\begin{align}
{t}_{\mu_1\ldots \mu_n} \nabla^{\mu_1}\cdots\nabla^{\mu_n} \phi &= {1\over n! \le({d-1\over 2}\ri)_n}{\cal T}(Y,K) (W\cdot \nabla)^n \Phi(Y)~,\cr
{t}_{\mu_1\ldots \mu_n} {f}^{\mu_1\ldots \mu_n}&= {1\over n! \le({d-1\over 2}\ri)_n}{\cal T}(Y,K) {\cal F}(Y,W)~.
\end{align}
where $t$ and $f$ are symmetric and traceless tensors. Note that for transverse polynomials, we have 
\be
\nabla \cdot K = K \cdot \nabla ~,
\ee
It is useful to notice that for polynomials of the form \eqref{eq:polenc} where the tensor is already symmetric, traceless and transverse, the projector reduces to $K=\left(\frac{d-1}{2}+n-1\right)\partial_{W}$. Since this will be the case in all our calculations, we will simply use $\partial_{W}$ to contract indices.

% If we want to project to objects that are just symmetric but not necessarily traceless, we would introduce
%  \begin{align}
%  \CT(Y;U)\equiv U^{A_{1}}\cdots U^{A_{n}}\CT_{A_{1}\cdots A_{n}}(Y)~,
%  \end{align}
% with the caveat the $U^2\neq 0$, but we still have $U\cdot Y=0$ and $\CT(Y, U+\alpha Y)=\CT(Y,U)$. The projection operator is 
% \be
% \hat K_A=\le({\partial\over \partial U^A}+Y^A Y\cdot {\partial\over \partial U}\ri)~,
% \ee
% and contractions are 
% \be
% {\hat t}_{\mu_1\ldots \mu_n} {\hat f}^{\mu_1\ldots \mu_n}= {1\over n! }{\cal T}(Y,\hat K) {\cal F}(Y,U)~.
% \ee
% with $\hat t$  and $\hat f$ are symmetric but not necessarily traceless vectors. 

\subsubsection{AdS$_{d+1}$ propagators}

 Here we follow \cite{Costa2014} and review some results of \cite{Sleight2016}; propagators in the AdS coordinates can be found in e.g. \cite{FreedmanMathurMatusisEtAl1999a,DHokerFreedman2002} among many other references. We are interested in describing the propagator of  a spin-$J$ field. In AdS coordinates, this field is a symmetric tensor that, in addition, satisfies the Fierz conditions
 \be
 \nabla^2 h_{\mu_1\ldots \mu_J}= M^2 h_{\mu_1\ldots \mu_J}~, \quad \nabla^{\mu_1} h_{\mu_1\ldots \mu_J}=0~, \quad h^\mu_{~\mu \mu_3\ldots \mu_J}=0~.
 \ee
These equations fully determine the AdS propagators, and the explicit answer are nicely casted in the embedding formalism. The  bulk--to--boundary propagator of a symmetric traceless field of rank $J$ can be written in a suggestive form 
 \begin{align}\label{eq:bdprop}
 G_{b\partial}^{\Delta|J}(Y_{j},P_{i};W_{j},Z_{i})=\mathcal{C}_{\Delta,J}\frac{\mathcal{H}_{ij}(Z_{i},W_{j})^{J}}{\Psi_{ij}^{\Delta}}~,
 \end{align}
 where $\mathcal{C}_{\Delta,J}$ is a normalization (which we will ignore), and
 \begin{align}\label{eq:psi}
 \Psi_{ij}\equiv -2 P_{i}\cdot Y_{j}~,\quad 
 \mathcal{H}_{ij}(Z_{i},W_{j})\equiv Z_{i}\cdot W_{j}+2 \frac{(W_{j}\cdot P_{i}) (Z_{i}\cdot Y_{j})}{\Psi_{ij}}~.
 \end{align}
 The mass squared is related to the conformal weight $\Delta$ of the dual operator as $M^2=\Delta(\Delta-d)-J$. This is the analogue of the CFT two point function \eqref{eq:2pcft}. It will be also useful to rewrite the bulk--to--boundary propagator as \cite{Sleight2016}
\be\label{eq:gbds}
G_{b\partial}^{\Delta|J}(Y,P;W,Z)={1\over (\Delta)_J}(\mathscr{D}_{P}(W,Z))^J \, G_{b\partial}^{\Delta|0}(Y,P)~,
\ee
 where
  \be
 \mathscr{D}_{P}(W,Z)= (Z \cdot W)\left( Z \cdot \frac{\partial}{\partial Z}- P \cdot \frac{\partial}{\partial P}\right)+(P\cdot W) \left(Z\cdot \frac{\partial}{\partial P}\right)~.
 \ee
 And it will also be convenient to cast the $n$-th derivative of $G_{b\partial}^{\Delta|J}$ in terms of scalar propagators:
 \begin{multline}\label{eq:id345}
 (W'\cdot \partial_Y)^n G_{b\partial}^{\Delta|J}(Y,P;W,Z) =  { 2^n \Gamma(\Delta+n)} \sum_{i=0}^J\sum_{k=0}^i {J\choose i}{i\choose k} {(n-k+1)_k \over  \Gamma(\Delta +i)} (W\cdot P)^i (W\cdot Z)^{J-i} \\ \times (W'\cdot Z)^k (W'\cdot P)^{n-k} (Z\cdot \partial_P)^{i-k}G_{b\partial}^{\Delta+n|0}(Y,P)~.
 \end{multline}
 
 The bulk--to--bulk propagator of a spin-$J$ fields can be written as\footnote{Note that \eqref{eq:bbprop} is not a homogeneous function of $Y$. In solving for the bulk-to-bulk operator the constrain $Y^2=-1$ is used, which breaks the homogeneity property of the polynomials in embedding space. }  
 \begin{align}\label{eq:bbprop}
 G_{bb}^{\Delta|J}(Y_{i},Y_{j};W_{i},W_{j})=\sum_{k=0}^{J}(W_{i}\cdot W_{j})^{J-k}(W_{i}\cdot Y_{j}W_{j}\cdot Y_{i})^{k}g_{k}(u)~,
 \end{align}
 where $ u=-1+Y_{ij}/2$ and $Y_{ij}\equiv -2 Y_{i}\cdot Y_{j}$. The functions $g_k$ can be written in terms of hypergeometric functions via 
 \begin{align}
 g_{k}(u)=\sum_{i=k}^{J}(-1)^{i+k}\left( \frac{i!}{j!} \right)^{2}\frac{h_{i}^{(k)}(u)}{(i-k)!}~,
 \end{align}
 where the recursion relation for $h_{i}$ is 
 \be
 h_k= c_k\Big((d-2k+2J-1)\le[(d+J-2)h_{k-1}+(1+u)h'_{k-1}\ri] + (2-k+J)h_{k-2}\Big)~,
 \ee
 where
 \be
 c_k=-{1+J-k\over k (d+2J-k-2)(\Delta + J -k-1)(d-\Delta+J-k-1)}~, 
 \ee
 and 
 \begin{align}
 h_{0}(u)=\frac{\Gamma(\Delta)}{2\pi^{h}\Gamma(\Delta+1-h)}(2u)^{-\Delta}{}_{2}F_{1}\left( \Delta,\Delta-h+\frac12,2\Delta-2h+1,-\frac{2}{u} \right).
 \end{align}

%%%%%%%%%%%%%%%%%%%%%%%%%%%%%%%%%%%%%%%%%%%%%%%%%%%%%%%%%%%%%%%%%%%%%
%%%%%%%%%%%%%%%%%%%%%%%%%%%%%%%%%%%%%%%%%%%%%%%%%%%%%%%%%%%%%%%%%%%%%
\section{Geodesic Witten diagrams}\label{sec:maingwd}

\begin{figure}
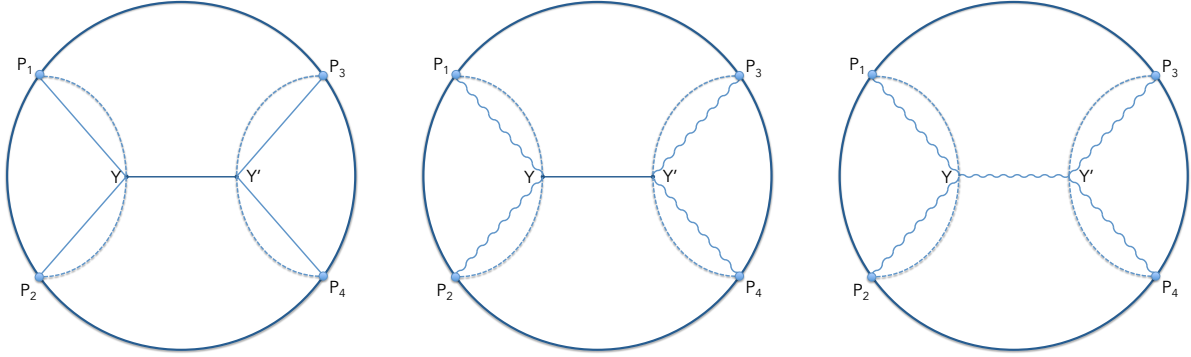

%\centering
{ \includegraphics[width=0.33\textwidth,page=9]{diagrams.pdf}
   \includegraphics[width=0.33\textwidth,page=10]{diagrams.pdf}
  \includegraphics[width=0.33\textwidth,page=11]{diagrams.pdf}
 }
  \caption{Examples of geodesic Witten diagrams in AdS$_{d+1}$. The doted line indicates that we are projecting the propagators over a geodesic that connects the endpoints. Straight lines correspond to scalar fields, while wavy lines are symmetric traceless tensors of spin $J$. The first diagram corresponds to the scalar block in \eqref{eq:gwd1}. The middle diagram (with scalar propagator in the exchange) will be the focus of section \ref{sec:gwdsc} and the last diagram (with a spin-$J$ field exchanged) is the focus of section \ref{sec:gwdspin}. }\label{fig:gwd}
\end{figure}

The idea placed forward in \cite{Hijano2016} was to consider the following object in AdS$_{d+1}$:
\begin{multline}\label{eq:gwd1}
{\cal W}_{\Delta | 0}(x_1,x_2,x_3,x_4)=\\ \int_{\gamma_{12}} d\lambda \int_{\gamma_{34}} d\lambda' G_{b\partial}^{\Delta_1|0}( y(\lambda),x_1)G_{b\partial}^{\Delta_2|0}(y(\lambda),x_2) G_{bb}^{\Delta|0}(y(\lambda),y'(\lambda'))G_{b\partial}^{\Delta_3|0}(y'(\lambda'),x_3)G_{b\partial}^{\Delta_4|0}(y'(\lambda'),x_4)~.
\end{multline}
Here $\gamma_{ij}$ is a geodesic that connects the boundary points $(x_i,x_j)$; $\lambda$ is an affine parameter for  $\gamma_{12}$ and $\lambda'$  for  $\gamma_{34}$. This is the simplest version of a {\it geodesic Witten diagram}: the expression involves bulk--to--boundary and bulk--to--bulk propagators in AdS projected along geodesics connecting the endpoints, as depicted in Fig. \ref{fig:gwd}. It was shown explicitly in \cite{Hijano2016} that ${\cal W}_{\Delta | 0}(x_1,x_2,x_3,x_4)$ gives the scalar conformal partial wave ${W}_{\Delta | 0}(x_1,x_2,x_3,x_4)$ as defined in \eqref{eq:cpw}, and there is evidence that it works correctly as we consider more general partial waves\cite{Hijano2016,Nishida2016}.

Our interest here is to explore cases where the external and internal lines have non-trivial spin.  In this section we will  give a prescription on how to obtain $W_{\Delta | l}^{l_1,l_2,l_3,l_4}(x_1,x_2,x_3,x_4)$  by using a basis of AdS$_{d+1}$ differential operators which will act on \eqref{eq:gwd1}. This should be viewed as the gravitational version of the relations in \eqref{eq:cpw}, where suitable tensor structures are built a by a set derivatives acting on $x_i$.  We stress that we will not use local cubic interactions to capture  the conformal partial wave in this section. We postpone to  section \ref{sec:interactions} the interpretation of this construction in terms of cubic interactions in the bulk.  
 
 \subsection{Construction of bulk differential operators: scalar exchanges}\label{sec:gwdsc}
 
To start we want to give an AdS analog of the CFT operators that generate tensor structures in spinning conformal partial waves. We recall that there are two class of operators
\be\label{eq:dh}
D_{i\, jk} ~,\qquad {\rm and} \qquad H_{ij}~.
\ee
The operators $D_{i\, jk}$, defined in \eqref{eq:dij}, are differential operators that basically raise spin at position $j$; these operators  we will map to  differential operators acting on bulk coordinates. $H_{ij}$, defined in \eqref{eq:VH}, raises the spin at position $i$ and $j$; it is not a differential operator, so its action will remain unchanged.  $H_{ij}$ does induce a cubic interaction and we will discuss its effect in section \ref{sec:interactions}.
  
\begin{figure}
\centering
{ \includegraphics[width=0.4\textwidth,page=6]{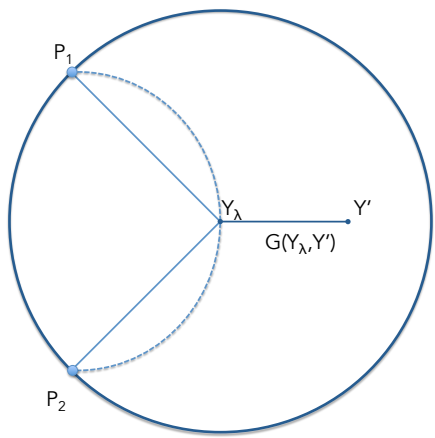}
 }
  \caption{A precursor diagram where two legs are in the boundary and one in the bulk. This type of object appears at intermediate steps when evaluating  conformal blocks.  }\label{gw1}
\end{figure}

The action of a single operator in \eqref{eq:dh}  on a conformal partial wave $W_{\Delta|l}(P_i)$  will affect either the pair $(P_1,P_2)$ or $(P_3,P_4)$, but not all points simultaneously. So let's consider the components in the integral \eqref{eq:gwd1} that only depends on $\gamma_{12}$ which connects $(P_1,P_2)$:
 \be\label{eq:w12}
  \int_{\gamma_{12}} d\lambda \, G_{b\partial}^{\Delta_1|0}( Y_\lambda,P_1)G_{b\partial}^{\Delta_2|0}( Y_\lambda,P_2) G_{bb}^{\Delta|0}(Y_\lambda,Y')~,
 \ee
 where we casted the propagators in embedding space.\footnote{We recall our notation: $Y^A$ denotes AdS points and $W^A$ are the auxiliary vectors that soak up bulk spin. The analogous CFT quantities are $P^A$ and $Z^A$, respectively.} Fig. \ref{gw1} depicts diagramatically the content in \eqref{eq:w12}, and we note that $Y'$ is not necessarily projected over $\gamma_{34}$.
 Here  $G_{b\partial}^{\Delta_1|0}(Y,P_1)\equiv G_{b\partial}^{\Delta_1|0}(Y,P_{1};0,0) $ given in \eqref{eq:bdprop}; in general we will omit dependence on variables that are not crucial for the equation in hand. 
 
  Using  Poincare coordinates, a geodesic that connects  $x_i$ with $x_j$ is 
\begin{align}\label{eq:geodefPoin}
\gamma_{ij}:\quad y^{\mu}(\lambda)=&(r(\lambda),{x}^{a}(\lambda))=\left( \frac{(x_{ij}^{2})^{\frac12}}{2 \cosh(\lambda)},
\frac{x_{i}^{a}+x_{j}^{a}}{2}+\frac{(x_{ij})^{a}}{2}\tanh(\lambda)\right)~,\quad x_{ij}\equiv x_i -x_j~,
\end{align}
and passing this information to the embedding formalism, we have
 \begin{align}\label{eq:geodef}
\gamma_{ij}:\qquad Y^{A}_\lambda\equiv \frac{e^{-\lambda}P_{i}^{A}+e^{\lambda}P_{j}^{A}}{\sqrt{P_{ij}}}~,\quad P_{ij}= -2P_i\cdot P_j~,
 \end{align}
 where we used \eqref{eq:px} and \eqref{eq:defny}. Evaluating \eqref{eq:w12} along $\gamma_{12}$ gives  
 \be\label{eq:w12P}
  {1\over ({P_{12}})^{(\Delta_1+\Delta_2)/2}}\int_{-\infty}^{\infty} d\lambda \, e^{-\Delta_{12}\lambda}\, G_{bb}^{\Delta|0}(Y_\lambda,Y')~, \qquad \Delta_{12}=\Delta_1-\Delta_2~.
 \ee
To increase the spin at $P_1$ and/or $P_2$ we would act on  \eqref{eq:w12P} with a combination of the differential operators in \eqref{eq:dij}. By inspection of the integral in  \eqref{eq:w12P},   $D_{i\, jk}$ has only a non-trivial action over the bulk--to--bulk propagators: $G_{b\partial}$ plays no role in building the OPE structures. Another way of staying this is to note that
\be
D_{k\, ij} G_{b\partial}^{\Delta_{n}|0}(Y_\lambda,P_{n})=0~,\qquad n=1,2~.
\ee
Hence,  the task ahead is to build a bulk differential operator that acts on the third leg of the diagram: $G_{bb}^{\Delta|0}(Y_\lambda,Y')$. 
 
Let's consider then a general function $G(Y_{\lambda }\cdot Y')$ that doesn't depend explicitly on $P_{i}$ (only through the geodesics in $Y_\lambda$), and further more with no $W$ dependence. We want to find differential operators $\mathcal{D}$ such that
\begin{align}\label{eq:defdopsbulkonf}
D_{k\, ij} G\left(Y_{\lambda }\cdot Y'\right) 
=\mathcal{D}_{k\, ij}G\left(Y_{\lambda }\cdot Y'\right)~,
\end{align}
where $\mathcal{D}_{k\, ij}$ has derivatives with respect to $Y'$ only. This equality implies that $\mathcal{D}$ has to satisfy the same basic properties those in $D$, listed in section \ref{sec:cftcorr}. 
 %\begin{enumerate}
%\item It must raise the degree in $Z_1$ up to $l_1$ and $Z_2$ up to $l_2$.
%\item It must take terms $O(Z_{n}^2, P_{n}^2, Z_n\cdot P_n)$ to terms of the same kind: keep pure gauge terms as pure gauge.
%\item It must map transverse functions to themselves, i.e. it commutes with $P_i\cdot \partial_{Z_i}$.  
%\end{enumerate}
The set of differential operators that satisfy our requirements is
\begin{align}\label{eq:dopsbulk}
\mathcal{D}_{1\,ij}&= Z_{i}\cdot Y' \,P_{i}\cdot \partial_{Y'}+\frac12 \Psi_{i Y'} \, Z_{i}\cdot \partial_{Y'}~,\notag\\
\mathcal{D}_{2\,ij}&=H_{ij}(Z_{i},Y')P_{j} \cdot\partial_{Y'} +\frac12 \Psi_{jY'}  H_{ij}(Z_{i},\partial_{Y'})~.
\end{align}
where $\Psi_{ij}$ is given in \eqref{eq:psi} and $H_{ij}(Z_i,Z_j)$ is defined in  \eqref{eq:VH}. The key property to constrain \eqref{eq:dopsbulk} is to demand transversality of the operators (i.e. that it commutes with $P_i\cdot \partial_{Z_i}$), and the rest follows from demanding \eqref{eq:defdopsbulkonf}. Note that these operators do not scale under $Y'\to \alpha Y'$, which leaves the homogeneity properties of the third field intact. $\mathcal{D}_{1\,ij}$ is increasing the spin by one and decreasing the dimension by one at position $i$, while $\mathcal{D}_{2\,ij}$ increases the spin at position $i$ by one and decreases the dimension by one at position $j$. The extra subscript $(1,2)$ in \eqref{eq:dopsbulk} is to keep the notation in the same line as in \eqref{eq:dij}.

To verify that  $\mathcal{D}$ has exactly the same effect as $D$, it is instructive to go through some identities. One can show the following relation by direct calculation
\begin{align}\label{eq:commrel}
[D_{k\,ij},\mathcal{D}_{k'\,i'j'}]f(Y')=[\mathcal{D}_{k\,ij},\mathcal{D}_{k'\,i'j'}]f(Y')~.
\end{align}
Let's call $D_{1}$, $D_{2}$ two generic operators of the form $D_{k\, ij}$, then
\begin{align}
D_{1}D_{2}(Y_{\lambda }\cdot Y')&=(D_{1}Y_{\lambda})\cdot (\mathcal{D}_{2}Y')+Y_{\lambda}\cdot (D_{1}\mathcal{D}_{2}Y')\notag\\
&=Y_{\lambda}\cdot (\mathcal{D}_{2}\mathcal{D}_{1}Y')+Y_\lambda\cdot ([D_{1},\mathcal{D}_{2}]Y')\notag\\
&=Y_\lambda \cdot (\mathcal{D}_{1}\mathcal{D}_{2}Y')
=\mathcal{D}_{1}\mathcal{D}_{2}(Y_{\lambda }\cdot Y')
\end{align}
where  in the third line we used  (\ref{eq:commrel}). Then for the product of an arbitrary number of operators,
\begin{align}\label{eq:dualopsinY2}
 D_{1}D_{2}\cdots D_{n}Y_{\lambda }\cdot Y'
 &=Y_{\lambda}\cdot (\mathcal{D}_{2}\cdots\mathcal{D}_{n}\mathcal{D}_{1}Y')+Y_{\lambda}\cdot (D_{1}\mathcal{D}_{2}\cdots\mathcal{D}_{n}Y')\notag\\
 &=Y_\lambda\cdot (\mathcal{D}_{1}\mathcal{D}_{2}\cdots\mathcal{D}_{n}Y')
 =\mathcal{D}_{1}\mathcal{D}_{2}\cdots\mathcal{D}_{n}Y_{\lambda }\cdot Y'
\end{align}
where in the first line we used the induction hypothesis for $n-1$ operators and in the second line we pushed $D_{1}$ through and used (\ref{eq:commrel}) to put everything in terms of $\mathcal{D}$. The conclusion is that boundary derivatives on geodesic integrals can be replaced by bulk derivatives:
\begin{align}\label{eq:dopmap}
&H_{12}^{n_{12}}(\mathcal{D}_{2,12}^{n_{1}}
\mathcal{D}_{2,21}^{n_{2}}\mathcal{D}_{1,12}^{m_{1}}\mathcal{D}_{1,21}^{m_{2}}
-D_{2,12}^{n_{1}}
D_{2,21}^{n_{2}}D_{1,12}^{m_{1}}D_{1,21}^{m_{2}})\notag\\ &\qquad \qquad\times
\int_{\gamma_{12}}d\lambda\,G_{b\partial}^{\Delta_1|0}(Y_{\lambda},P_{1})
G_{b\partial} ^{\Delta_2|0}(Y_{\lambda},P_{2})G_{bb}^{\Delta|0}(Y_{\lambda},Y')
=0~.
\end{align}

%Notice that $\mathcal{D}$ in (\ref{eq:dopsbulk}) is invariant under $\partial_{Y_{3}}\to \partial_{Y_{3}}+Y_{3}(Y_{3}\cdot \partial_{Y_{3}})$, and there is no $W$ dependence, we can therefore replace $\partial_{Y_{3}}$ by the covariant derivative $\nabla_{Y_{3}}$, defined as 
%\begin{align}
%\nabla_{Y_{3}\,A}=\frac{\partial}{\partial Y_3^{A}}+Y_{3\, A}\left( Y_3\cdot \frac{\partial}{\partial Y_{3}} \right)+W_{3\,A}\left( Y_{3}\cdot \frac{\partial }{\partial W_{3}} \right).
%\end{align}

We just found that the dual of $D$ are derivatives with respect to $Y'$. However, the generic form of  this differential operators is $\mathcal{D}(Y')=Y'^{A}Z_{i}^{B}S_{ABC}\partial_{Y'}^{C}$, where $S$ is antisymmetric under $A\leftrightarrow C$ due to \eqref{eq:dopsbulk}. Hence 
\begin{align}\label{eq:Santisymm}
& Y'^{A}Z_{i}^{B}S_{ABC}\partial_{Y'}^{C}Y_{\lambda }\cdot Y'=-Y_{\lambda}^{A}Z_{i}^{B}S_{ABC}\partial_{Y_{\lambda}}^{C}Y_{\lambda }\cdot Y'
\quad
 \Rightarrow \quad \mathcal{D}_{k\,ij}(Y')Y_{\lambda }\cdot Y'=-\mathcal{D}_{k\,ij}(Y_{\lambda})Y_{\lambda }\cdot Y'~.
 \end{align} 
Using (\ref{eq:Santisymm}) it is easy to show that for more derivatives, 
\begin{align}\label{eq:DY2toDY1}
\mathcal{D}_{k_{1}\,i_{1}j_{1}}(Y')\cdots \mathcal{D}_{k_{n}\,i_{n}j_{n}}(Y')Y_{\lambda }\cdot Y'
=(-1)^{n}\mathcal{D}_{k_{n}\,i_{n}j_{n}}(Y_{\lambda})\cdots \mathcal{D}_{k_{1}\,i_{1}j_{1}}(Y_{\lambda})Y_{\lambda }\cdot Y'~.
\end{align}
This of course also holds when the derivatives act on $G(Y_{\lambda }\cdot Y')$. It is interesting to note that the action of ${\cal D}(Y_\lambda)$ on bulk--to--boundary operators is trivial, i.e.
\be\label{eq:dg}
\mathcal{D}_{k\, i j}(Y_{\lambda}) G_{b\partial}^{\Delta_{1,2}|0}( Y_\lambda,P_{1,2}) =0~.
\ee
However,
\be\label{eq:dg1}
\mathcal{D}_{k'\, i' j'}\cdots \mathcal{D}_{k\, i j}(Y_{\lambda}) G_{b\partial}^{\Delta_{1,2}|0}( Y_\lambda,P_{1,2}) \neq0~,
\ee
because \eqref{eq:dg} relies on properties of the geodesic $\gamma_{12}$, and in \eqref{eq:dg1} the operation of taking derivatives with respect to $Y$ does not commute with projecting on $\gamma_{12}$\footnote{For $\mathcal{D}_{1\,21}$ and $\mathcal{D}_{2\,12}$, \eqref{eq:dg} is true without projecting on $\gamma_{12}$. Furthermore, \eqref{eq:dg1} is true only if the $\mathcal{D}$'s do not commute. However, we will use \eqref{eq:dopmap1} to treat all the $\mathcal{D}$'s in the same footing.}. Hence, as we generate tensorial structures using  ${\cal D}(Y_\lambda)$, it only acts on $G_{bb}$, i.e.
\begin{align}\label{eq:dopmap1}
& (-1)^{N}\int_{\gamma_{12}}d\lambda\,G_{b\partial}^{\Delta_1|0}(Y_{\lambda},P_{1})
G_{b\partial} ^{\Delta_2|0}(Y_{\lambda},P_{2})\mathcal{D}_{1,21}^{m_{2}} \mathcal{D}_{1,12}^{m_{1}}\mathcal{D}_{2,21}^{n_{2}}\mathcal{D}_{2,12}^{n_{1}}G_{bb}^{\Delta_3|0}(Y_{\lambda},Y') =\cr
& D_{2,12}^{n_{1}}D_{2,21}^{n_{2}}D_{1,12}^{m_{1}}D_{1,21}^{m_{2}}\int_{\gamma_{12}}d\lambda\,G_{b\partial}^{\Delta_1|0}(Y_{\lambda},P_{1})
G_{b\partial} ^{\Delta_2|0}(Y_{\lambda},P_{2})G_{bb}^{\Delta_3|0}(Y_{\lambda},Y')~,
\end{align}
where $N\equiv m_{1}+m_{2}+n_{1}+n_{2}$.

From here we see how to cast conformal partial waves where the exchanged field is a scalar field (dual to a scalar primary  $\cal O$ of conformal dimension $\Delta$): the version of  \eqref{eq:cpw}  in gravitational language is
\begin{align}\label{eq:grcpw}
W_{\Delta | 0}^{l_1,l_2,l_3,l_4}(P_i;Z_i)= {\cal W}_{\Delta | 0}[\mathcal{D}_{\rm left}(Y_\lambda),\mathcal{D}_{\rm right}(Y'_{\lambda'})]~,
\end{align}
where 
\begin{multline}
{\cal W}_{\Delta | 0}[\mathcal{D}_{\rm left}(Y_\lambda),\mathcal{D}_{\rm right}(Y'_{\lambda'})]\equiv\\ \int_{\gamma_{12}} \int_{\gamma_{34}} \,G_{b\partial}^{\Delta_1|0}(Y_{\lambda},P_{1})G_{b\partial} ^{\Delta_2|0}(Y_{\lambda},P_{2})
\left[\mathcal{D}_{\rm left}(Y_\lambda)\mathcal{D}_{\rm right}(Y'_{\lambda'})G_{bb}^{\Delta|0}(Y_{\lambda},Y'_{\lambda'})  \right] G_{b\partial}^{\Delta_3|0}(P_3, Y'_{\lambda'})G_{b\partial}^{\Delta_4|0}(P_4, Y'_{\lambda'})~.
\end{multline} 

To close this discussion, we record another convenient way to re-write \eqref{eq:dopsbulk}:
\begin{align}\label{eq:dopsbulkcov}
\mathcal{D}_{1\,ij}(Y_\lambda)&= \frac{\Psi_{i\lambda}}{2}\mathcal{H}_{i\lambda}(Z_{i},\partial_{Y_\lambda})~,\notag\\
\mathcal{D}_{2\,ij}(Y_\lambda)&=\frac{\Psi_{j\lambda}}{2}\left[\mathcal{H}_{i\lambda}(Z_{i},\partial_{Y_\lambda})+2 \mathcal{V}_{\partial\,i,j\lambda}(Z_{i})\mathcal{V}_{b\,\lambda,ij}(\partial_{Y_\lambda})\right]~,
\end{align}
where $\mathcal{H}_{ij}$ is given in \eqref{eq:psi}, and we defined 
\begin{align}
\mathcal{V}_{\partial\,i,jm}(Z_{i})&=\frac{\Psi_{im}Z_{i}\cdot P_{j}-P_{ij}Z_{i}\cdot Y_{m}}{\sqrt{\Psi_{im}\Psi_{jm}P_{ij}}}~,\\
\mathcal{V}_{b\,m,ij}(W_{m})&=\frac{\Psi_{jm}W_{m}\cdot P_{i}-\Psi_{im}W_{m}\cdot P_{j}}{\sqrt{\Psi_{im}\Psi_{jm}P_{ij}}}~,
\end{align}
which can be viewed as the analogous CFT in \eqref{eq:VH}.

%%%%%%%%%%%%%%%%%%%%%%%%%%%%%%%%%%%%%%%%%%%%%%%%%%%%%%%%%%%%%%%%%%%%%
%%%%%%%%%%%%%%%%%%%%%%%%%%%%%%%%%%%%%%%%%%%%%%%%%%%%%%%%%%%%%%%%%%%%%

\subsection{Construction of bulk differential operators: spin exchanges}\label{sec:gwdspin}

We now generalize the discussion to include spin fields in the exchange diagram. The prescription given in  \cite{Hijano2016} for spinning exchanged operators is that the bulk--to--bulk propagator for the spin $J$ field is contracted with the velocities of $Y_{\lambda}$ and $Y'_{\lambda'}$, i.e.
\begin{align}\label{eq:bbpropspin}
G_{bb}^{\Delta|J}(Y_{\lambda},Y'_{\lambda'}) \equiv G_{bb}^{\Delta|J}\left(Y_{\lambda},Y'_{\lambda'};\frac{d Y_{\lambda}}{d \lambda},\frac{d Y'_{\lambda'}}{d\lambda'}\right)~.
\end{align}
This corresponds to the pullback of the propagator \eqref{eq:bbprop} along both geodesics in the diagram. Hence, a geodesic diagram that evaluates the conformal partial wave with a spin exchange is
\begin{multline}\label{eq:gwd2}
{\cal W}_{\Delta | J}(P_1,P_2,P_3,P_4)=\\ \int_{\gamma_{12}} \int_{\gamma_{34}} G_{b\partial}^{\Delta_1|0}( Y_\lambda,P_1)G_{b\partial}^{\Delta_2|0}(Y_\lambda,P_2) G_{bb}^{\Delta|J}(Y_{\lambda},Y'_{\lambda'})G_{b\partial}^{\Delta_3|0}(Y'_{\lambda'},P_3)G_{b\partial}^{\Delta_4|0}(Y'_{\lambda'},P_4)~.
\end{multline}

In manipulating \eqref{eq:bbpropspin} to increase the spin of the external legs, we need to treat the contractions with $\frac{d Y_{\lambda}}{d \lambda}$ with some care. First, it is important  to note that $D_{k\, ij}$ commutes with $\frac{d}{d \lambda}$, and hence its action on $G_{bb}^{\Delta|J}(Y_{\lambda},Y'_{\lambda'})$ in \eqref{eq:gwd2} is straightforward. However, we need to establish how $\mathcal{D}_{k\, ij}$ acts \eqref{eq:bbpropspin}, and this requires understanding how to cast $\frac{d}{d \lambda}$ as a covariant operation. It is easy to check by direct computation that this can be done in two ways:
\begin{align}\label{eq:dldy}
\frac{d}{d \lambda} = -2 P_{12}^{-1}\Psi_{2 \lambda} P_{1}\cdot \nabla_{Y_{\lambda}} = 2 P_{12}^{-1}\Psi_{1 \lambda} P_{2}\cdot \nabla_{Y_{\lambda}}~.
\end{align}
But the commutator of $\mathcal{D}_{k\, ij}$ with $\frac{d}{d \lambda}$ will depend on which equality we use. For example 
\begin{align}
D_{1\,12}\frac{d Y_{\lambda}}{d \lambda}&=-\mathcal{D}_{1\, 12}(Y_{\lambda})(-2 P_{12}^{-1}\Psi_{2 \lambda} P_{1}\cdot \nabla_{Y_{\lambda}})Y_{\lambda}~,\\
D_{2\,21}\frac{d Y_{\lambda}}{d \lambda}&=-\mathcal{D}_{2\, 21}(Y_{\lambda})(-2 P_{12}^{-1}\Psi_{2 \lambda} P_{1}\cdot \nabla_{Y_{\lambda}})Y_{\lambda}~,
\end{align}
which is the expected result by (\ref{eq:dualopsinY2}) and (\ref{eq:DY2toDY1}). Unfortunately, the two other $D$'s have the wrong sign relative to  (\ref{eq:dualopsinY2}) and (\ref{eq:DY2toDY1}): 
\begin{align}
D_{1\,21}\frac{d Y_{\lambda}}{d \lambda}&=\mathcal{D}_{1\, 21}(Y_{\lambda})(-2 P_{12}^{-1}\Psi_{2 \lambda} P_{1}\cdot \nabla_{Y_{\lambda}})Y_{\lambda}~,\\
D_{2\,12}\frac{d Y_{\lambda}}{d \lambda}&=\mathcal{D}_{2\, 12}(Y_{\lambda})(-2 P_{12}^{-1}\Psi_{2 \lambda} P_{1}\cdot \nabla_{Y_{\lambda}})Y_{\lambda}~.
\end{align}
Using the other implementation of $\frac{d}{d \lambda}$ alternates the signs. In order to avoid this implementation problem, we formally define 
\begin{align}\label{eq:mcDcomm}
\left[ \mathcal{D}_{k\, ij}(Y_{\lambda}), \frac{d}{d \lambda} \right] \equiv 0~.
\end{align}
This implies that as we encounter quantities that contain explicit derivatives of $\lambda$ we will manipulate them by first acting with $\mathcal{D}_{k\, ij}(Y_{\lambda})$ and then taking the derivative with respect to $\lambda$. For instance,
\begin{align}
D_{k\, ij}\frac{d Y_{\lambda}}{d \lambda}\cdot \frac{d Y'_{\lambda'}}{d \lambda'}&=\frac{d}{d \lambda} \frac{d}{d \lambda'} D_{k\,ij}Y_{\lambda}\cdot Y'_{\lambda'}\cr
&=-\frac{d}{d \lambda} \frac{d}{d \lambda'} \mathcal{D}_{k\,ij}(Y_{\lambda})Y_{\lambda}\cdot Y'_{\lambda'}~.
\end{align}
 Given this implementation of the differential operators,  the partial wave in gravitational language (\ref{eq:grcpw}) generalizes to spinning exchanges by using (\ref{eq:bbpropspin}) and (\ref{eq:mcDcomm}). This shows that for each partial wave $W_{\Delta|J}^{l_1,l_2,l_3,l_4}(P_i;Z_i)$ in the boundary CFT there is a counterpart geodesic integral in AdS  ${\cal W}_{\Delta|J}^{l_1,l_2,l_3,l_4}(P_i;Z_i)$ that reproduces the same quantity. 

%%%%%%%%%%%%%%%%%%%%%%%%%%%%%%%%%%%%%%%%%%%%%%%%%%%%%%%%%%%%%%%%%%%%%
%%%%%%%%%%%%%%%%%%%%%%%%%%%%%%%%%%%%%%%%%%%%%%%%%%%%%%%%%%%%%%%%%%%%%
\section{Identification of gravitational interactions via geodesic diagrams}\label{sec:interactions}

We have given in the previous section a systematic procedure to build the appropriate tensor structures $V_{i,jk}$ and $H_{ij}$ appearing in conformal partial waves by using directly bulk differential operators ${\cal D}_{i\, jk}(Y_\lambda)$. Using this method, we would like to identify the gravitational interactions that the operators  ${\cal D}_{i\, jk}(Y_\lambda)$ are capturing.

The identification of tensor structures with gravitational interactions has been successfully carried out in \cite{Sleight2016b}:  all possible cubic vertices in AdS$_{d+1}$ where mapped to the tensor structures of a CFT$_{d}$ via Witten diagrams for three point functions.  Here we would like to revisit this identification using instead as a building block diagrams in AdS that are projected over geodesic integrals rather than volume integrals; and as we will show below, the geodesic diagrams do suffer from some non-trivial cancellations for certain derivative interactions. 
%that generate some tension with the map developed in \cite{Sleight2016b} and the proposal in \cite{Hijano2016}.

\begin{figure}
%\centering
{ \includegraphics[width=0.33\textwidth,page=1]{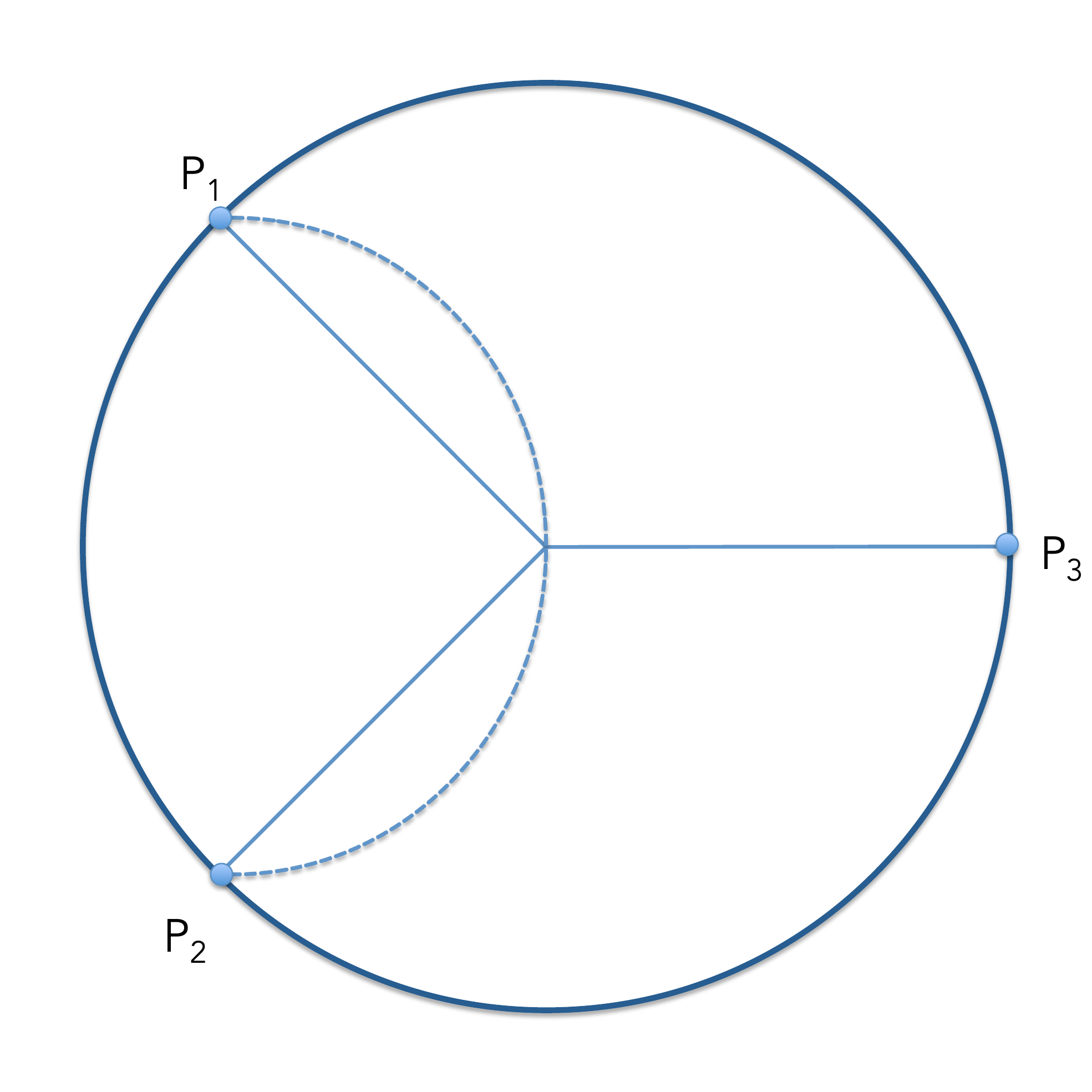}
   \includegraphics[width=0.33\textwidth,page=2]{diagrams.pdf}
  \includegraphics[width=0.33\textwidth,page=3]{diagrams.pdf}
 }
  \caption{Examples of geodesic Witten diagrams in AdS$_{d+1}$ that capture three point functions. Straight lines correspond to scalar propagators, while wavy lines denote symmetric traceless spin-$J$ fields; $P_i$ is the boundary position in embedding formalism. The dotted line denotes the geodesic over which we integrate. Note that the second and third diagram only differ by the choice of geodesic.     }\label{gwd3}
\end{figure}

For the discussion in this section it is sufficient to consider the following object 
\be\label{eq:w3}
\int_{\gamma_{ij}} d\lambda \, G_{b\partial}^{\Delta_1|0}( y(\lambda),x_1)G_{b\partial}^{\Delta_2|0}(y(\lambda),x_2) G_{b\partial}^{\Delta_3|0}( y(\lambda),x_3)~.
\ee
Here $\gamma_{ij}$ is a geodesic that connects a pair of endpoints $(x_i,x_j)$. Rather interestingly, it was noted in \cite{Nishida2016} that  this integral actually reproduces the CFT three point function for scalar primaries; this equivalence is regardless the choice of endpoints, with different choices just giving different numerical factors.\footnote{The results in \cite{Czech2016a,Boer2016} as well suggested  that \eqref{eq:w3} reproduces correlation functions of three scalar primaries.}     The type of diagrams we will be considering are depicted in Fig. \ref{gwd3}, where the dotted line represents which geodesic we will integrate over. We will first attempt to rebuild interactions using these geodesic integrals, and at the end of this section we will contrast with the results in \cite{Sleight2016b}.

\subsection{Sampling three point functions via geodesics diagrams}\label{sec:int}

In this subsection we will go through some explicit computations of three point functions using the method developed in section \ref{sec:gwdsc}. Our goal is not to check that our bulk results match with the CFT values (which they do); our goal is to illustrate how these operators ${\cal D}_{i\, jk}(Y_\lambda)$, and hence $(V_{i,jk},H_{ij})$, map up to local AdS interactions. 

%Even for a fixed number of fields, the set of possible local AdS interactions it is obviously larger than the set of tensorial structures that define each OPE structure. 
%As we carryout the computation, we will see that the homogeneity properties of the bulk-to-boundary operators introduces some ambiguity on how to re-cast a given integrand in the bulk Witten diagram: multiple derivatives acting on the fields will usually just modify the overall coefficient of the diagram, but not its tensorial structure.  Our aim will be to identify the least number of interactions with minimal number of derivatives that give rise to each OPE structure. 

Our seed to all further computation is the three point function of three scalar primaries. In terms of geodesic integrals, we can write the scalar three-point function in the boundary as 
\begin{align}\label{eq:TPFdef}
 T(\Delta_{1},\Delta_{2},\Delta_{3}) &=c_{\Delta_{1}\Delta_{2}\Delta_{3}}
 \int_{\gamma_{12}} d\lambda\,
 G_{b \partial}^{\Delta_{1}|0}(Y_\lambda,P_{1})
 G_{b \partial}^{\Delta_{2}|0}(Y_\lambda,P_{2})
 G_{b \partial}^{\Delta_{3}|0}(Y_\lambda,P_3)\cr&=
 {1\over (P_{12})^{(\Delta_1+\Delta_2 -\Delta_3)/2} (P_{23})^{(\Delta_2+\Delta_3 -\Delta_1)/2}(P_{13})^{(\Delta_1+\Delta_3 -\Delta_2)/2}  }
~,
 \end{align} 
 where 
 \begin{align}\label{factor3point}
 c_{\Delta_{1}\Delta_{2}\Delta_{3}}=
 \frac{2 \Gamma(\Delta_{3})}{\Gamma \left( \frac{-\Delta_{1}+\Delta_{2}+\Delta_{3}}{2} \right)\Gamma \left( \frac{\Delta_{1}-\Delta_{2}+\Delta_{3}}{2} \right) }~.
 \end{align}
Here we are ignoring  the normalization of $G_{b \partial}$ in \eqref{eq:bdprop} and the gamma functions in $c_{\Delta_{1}\Delta_{2}\Delta_{3}}$ result from the integration over the geodesic $\gamma_{12}$. $G_{\Delta_1,\Delta_2,\Delta_3|0,0,0}=T(\Delta_{1},\Delta_{2},\Delta_{3})$ is the CFT$_d$ three point function in \eqref{eq:3scalars} casted as a geodesic integral in AdS$_{d+1}$.   %The integrand evaluated at the geodesic is 
%\begin{align}
%e^{\lambda(-\Delta_{1}+\Delta_{2}+\Delta_{3})}P_{12}^{\frac12(-\Delta_{1}-\Delta_{2}+\Delta_{3})}(P_{13}+e^{2 \lambda}P_{23})^{-\Delta_{3}}~.
%\end{align}
 
\subsubsection{Example: Vector-scalar-scalar}\label{sec:vss}
To start, we consider the three point function of one vector and two scalar operators as built from scalar operators. Following the CFT discussion in section \ref{sec:cftcorr}, in this case there is only one tensor structure which can be written in two ways:
\begin{align}\label{eq:v123}
G_{\Delta_1,\Delta_2,\Delta_3|1,0,0}&=V_{1,23}T(\Delta_{1},\Delta_{2},\Delta_{3})\cr
&=\frac{2D_{1\,12}}{-1- \Delta_{1}+\Delta_{2}+\Delta_{3} }T(\Delta_{1}+1,\Delta_{2},\Delta_{3})
\notag\\&=\frac{2D_{2\,12}}{-1+\Delta_{1}-\Delta_{2}+\Delta_{3} }T(\Delta_{1},\Delta_{2}+1,\Delta_{3})~.
\end{align}
We would like to extract which local bulk interaction can capture the left hand side of \eqref{eq:v123}. Let's choose the first equality for concreteness. Using \eqref{eq:dopmap1}, the bulk calculation is 
\begin{align}\label{eq:voo}
&\frac{2c_{\Delta_{1}+1\Delta_{2}\Delta_{3}}}{1+ \Delta_{1}-\Delta_{2}-\Delta_{3} }
 \int_{\gamma_{12}} d\lambda\,
 G_{b \partial}^{\Delta_1 +1|0}(Y_{\lambda},P_{1})
 G_{b \partial}^{\Delta_2|0}(Y_{\lambda},P_{2})
 \mathcal{D}_{1,12}(Y_\lambda)
 G_{b \partial}^{\Delta_3|0}(Y_{\lambda},P_{3})\cr
 &=
\frac{c_{\Delta_{1}+1\Delta_{2}\Delta_{3}}}{1+ \Delta_{1}-\Delta_{2}-\Delta_{3} }
 \int_{\gamma_{12}}d\lambda\,
 G_{b \partial}^{\Delta_1 +1|0}(Y_{\lambda},P_{1})
  G_{b \partial}^{\Delta_2|0}(Y_{\lambda},P_{2})
 \Psi_{1\lambda}\mathcal{H}_{11}(Z_{1}, \partial_{Y_{\lambda}})
 G_{b \partial}^{\Delta_3|0}(Y_{\lambda},P_{3})
 \cr
 &= \frac{c_{\Delta_{1}+1\Delta_{2}\Delta_{3}}}{1+ \Delta_{1}-\Delta_{2}-\Delta_{3} }
 \int_{\gamma_{12}}d\lambda\,
 G_{b \partial}^{\Delta_1|1}(Y_{\lambda},P_{1};\partial_W,Z_1)
 G_{b \partial}^{\Delta_2|0}(Y_{\lambda},P_{2})
(W\cdot\partial_{Y_\lambda})
 G_{b \partial}^{\Delta_3|0}(Y_{\lambda},P_{3})~.
\end{align}
The contraction appearing inside the integral  can be attributed to the following local AdS interaction
\be\label{eq:i1}
 A_1^{\mu}\phi_2\partial_{\mu}\phi_3 ~,
 \ee
 where $\phi_i$ is a bulk scalar of mass $M_i^2=\Delta_i(\Delta_i-d)$ and the massive vector $A_{1\mu}$ has $M_1^2=\Delta_1(\Delta_1-d)-1$. It is interesting to note that from this computation alone we could not infer that there is another potential interaction: $A_1^{\mu}\phi_3\partial_{\mu}\phi_2$.  This particular interaction is absent because $A_1^{\mu}\partial_{\mu}\phi_2$  vanishes when evaluated over the geodesic $\gamma_{12}$ due to \eqref{eq:dg}. However, it would have been the natural interaction if we instead perform the integral over  $\gamma_{13}$ in \eqref{eq:voo}. Hence a natural identification of the tensor structure in \eqref{eq:v123} with gravitational interactions is
 \be\label{eq:i1}
 V_{1,23}: \quad A_1^{\mu}\phi_2\partial_{\mu}\phi_3  \qquad {\rm and} \qquad A_1^{\mu}\phi_3\partial_{\mu}\phi_2~.
 \ee
If we used gauge invariance  we could constraint this combination to insist that $A_1$ couples to a conserved current (for us, however, the vector $A_1$ is massive). From the perspective of the usual Witten diagrams, which involve bulk integrals, these two interactions are indistinguishable up to normalizations, since they can be related after integrating by parts. In a geodesic diagram one has to take both into account; in our opinion, it is natural to expect that all pairings of endpoints $P_i$ have to reproduce the same tensor structure.    

%At this stage it is important to note that there is some ambiguity in mapping \eqref{eq:voo} to \eqref{eq:i1}. Since the bulk-to-boundary propagators are homogenous functions, we could add an arbitrary number of derivatives to \eqref{eq:i1} while reproducing the tensor structure in \eqref{eq:v123}. This is an ambiguity that later on we will use to our advantage, but as we mentioned before,  we will identify the OPE structures with gravitational interactions that involve the minimal number of derivatives on the fields.

\subsubsection{Example: Vector-vector-scalar}\label{sec:vvs} 
Moving on to the next level of complexity, we now consider the geodesic integral that would reproduce the three point function of two spin-1 fields and one scalar field. There are two tensor structures involved in this correlator, and  similar to the previous case, there are several combinations of derivatives that capture these structures. Choosing the combination in \eqref{eq:vv}, we have in CFT notation that one tensor structure is
\begin{align}\label{eq:vvs}
V_{1,23}V_{2,13}T(\Delta_{1},\Delta_{2},\Delta_{3})=
-\frac{4D_{1\,12}D_{1\,21}T(\Delta_{1}+1,\Delta_{2}+1,\Delta_{3})}{(\Delta_{1}-\Delta_{2})^{2}-\Delta_{3}^{2}}
+\frac{H_{12}T(\Delta_{1},\Delta_{2},\Delta_{3})}{-\Delta_{1}+\Delta_{2}+\Delta_{3}}~,
\end{align}
whereas the other tensor structure is simply 
\begin{align}\label{eq:th12}
H_{12}\,T(\Delta_{1},\Delta_{2},\Delta_{3})~.
\end{align}
$G_{\Delta_1,\Delta_2,\Delta_3|1,1,0}$ is the linear superposition of \eqref{eq:vvs} and \eqref{eq:th12}.

\begin{figure}
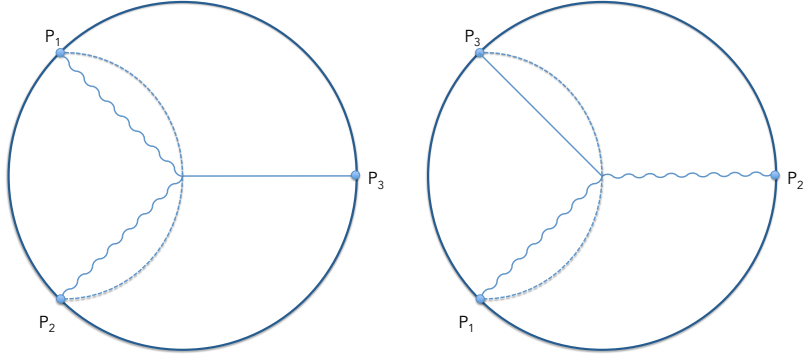

\centering
{ \includegraphics[width=0.33\textwidth,page=4]{diagrams.pdf}
   \includegraphics[width=0.33\textwidth,page=8]{diagrams.pdf}
 }
  \caption{ The diagrams here differ by the choice of geodesic. Depending on this choice, a given interaction will give rise to a different tensor structure.  }\label{fig:gwd3}
\end{figure}

As it was already hinted by our previous example, the identification of the interaction will depend on the geodesic we choose to integrate over. To start, let us consider casting  $T(\Delta_{1},\Delta_{2},\Delta_{3})$ exactly as in \eqref{eq:TPFdef}: the geodesic is $\gamma_{12}$ which connects at the positions with non-trivial spin (first diagram in Fig \ref{fig:gwd3}). For this choice of geodesic, the second tensor structure is straightforward to cast as a bulk interaction integrated over the geodesic. From the definitions \eqref{eq:VH} and \eqref{eq:psi}, one can show that
\be\label{eq:h123}
H_{12}={\cal H}_{1\lambda}(Z_1,\partial_W){\cal H}_{2\lambda}(Z_2,W) ~,
\ee
 where the right hand side is evaluated over the geodesic $\gamma_{12}$. Replacing this identity in \eqref{eq:th12}, we find
\begin{align}\label{eq:hint}
H_{12}\,T(\Delta_{1},\Delta_{2},\Delta_{3}) = c_{\Delta_{1},\Delta_{2},\Delta_{3}} \int_{\gamma_{12}} G_{b\partial}^{\Delta_1|1}(Y_\lambda,P_1;\partial_{W},Z_1)G_{b\partial}^{\Delta_2|1}(Y_\lambda,P_2;W,Z_2) G_{b\partial}^{\Delta_3|0}(Y_\lambda,P_3)~.
\end{align}
This contact term is simply in physical space the interaction
\be\label{eq:aap}
H_{12}:\qquad A_{1\mu}A^{\mu}_2 \phi_3~.
\ee
This contraction will be generic every time our tensorial structure involves $H_{12}$. In general we will have the following relation
\be\label{eq:aap1}
(H_{12})^n=\left({\cal H}_{1\lambda}(Z_1,\partial_{W}){\cal H}_{2\lambda}(Z_2,W)\ri)^n: \quad  \quad h_{1\mu_1\ldots \mu_n}h^{\mu_1\ldots \mu_n}_2 \phi_3~,
\ee
where $(H_{12})^n$ generates one of the tensor structure for a tensor-tensor-scalar three point function, and the natural  bulk interaction is the contraction of symmetric traceless tensors coupled minimally with a scalar.

For the other tensor structure, a bit more work is required. Let's first manipulate the first term in \eqref{eq:vvs}; using \eqref{eq:DY2toDY1} we can write
\begin{align}\label{eq:dd11}
D_{1\,12}D_{1\,21} G_{b\partial}^{\Delta_3|0}(Y_\lambda,P_3) &=\mathcal{D}_{1\,21}(Y_{\lambda})\mathcal{D}_{1\,12}(Y_{\lambda}) G_{b\partial}^{\Delta_3|0}(Y_\lambda,P_3)
\cr&=\frac18 \Psi_{1\lambda}\Psi_{2\lambda}\mathcal{H}_{1\lambda}(Z_{1},\partial_W)\mathcal{H}_{2\lambda}(Z_{2},\partial_W)(W\cdot \partial_{Y_{\lambda}})^2 G_{b\partial}^{\Delta_3|0}(Y_\lambda,P_3) \cr &
\quad +\frac12 H_{12}\Psi_{2\lambda}P_{1}\cdot \partial_{Y_{\lambda}} G_{b\partial}^{\Delta_3|0}(Y_\lambda,P_3)~.
\end{align}
Applying this expression to \eqref{eq:vvs} gives\footnote{The fastest way to reproduce \eqref{eq:v1} from \eqref{eq:dd11} is by using the explicit form of $G_{b\partial}^{\Delta_3|0}(Y_\lambda,P_3)$. An alternative route, which is more general, is to use \eqref{eq:dldy}: from here we can integrate by parts and rearrange the terms appropriately. This second route allows us to use \eqref{eq:VVOstr} when at the third leg of the vertex we have bulk--to--bulk propagators rather than bulk--to--boundary.  } 
\begin{align}\label{eq:v1}
&-\frac{4D_{1\,12}D_{1\,21}}{(\Delta_{1}-\Delta_{2})^{2}-\Delta_{3}^{2}}T(\Delta_{1}+1,\Delta_{2}+1,\Delta_{3})=-{4 c_{\Delta_1+1,\Delta_2+1,\Delta_3}\over (\Delta_{1}-\Delta_{2})^{2}-\Delta_{3}^{2}}\int_{\gamma_{12}}
 G_{b \partial}^{\Delta_{1}+1|0}
 G_{b \partial}^{\Delta_{2}+1|0}
\mathcal{D}_{1,21}\mathcal{D}_{1,12}G_{b \partial}^{\Delta_{3}|0}
\cr
&\qquad\qquad\qquad=-\frac12{c_{\Delta_1+1,\Delta_2+1,\Delta_3}\over (\Delta_{1}-\Delta_{2})^{2}-\Delta_{3}^{2}} \int_{\gamma_{12}}  G_{b \partial}^{\Delta_{1}|1}(\partial_W)
 G_{b \partial}^{\Delta_{2}|1}(\partial_W)
(W\cdot \partial_{Y_{\lambda}})^2G_{b \partial}^{\Delta_{3}|0}\cr&\qquad\qquad\qquad\qquad
-{1\over -\Delta_{1}+\Delta_{2}+\Delta_{3}}H_{12} T(\Delta_1,\Delta_2,\Delta_3)~.
\end{align}
Replacing \eqref{eq:v1} in \eqref{eq:vvs} results in
\begin{multline}\label{eq:VVOstr}
V_{1,23}V_{2,13}T(\Delta_{1},\Delta_{2},\Delta_{3})\\
 =-\frac{c_{\Delta_{1}+1 \Delta_{2}+1 \Delta_{3}}}{2 ((\Delta_{1}-\Delta_{2})^{2}-\Delta_{3}^{2})}\int_{\gamma_{12}}  G_{b \partial}^{\Delta_{1}|1}(\partial_W)
 G_{b \partial}^{\Delta_{2}|1}(\partial_W)
(W\cdot \partial_{Y_{\lambda}})^2G_{b \partial}^{\Delta_{3}|0}~.
%\\
%= -\frac{c_{\Delta_{1}+1 \Delta_{2}+1 \Delta_{3}}}{2 ((\Delta_{1}-\Delta_{2})^{2}-\Delta_{3}^{2})}\int_{\gamma_{12}} d\lambda  G_{b \partial}^{\Delta_{1}|1}(\partial_W)
 %G_{b \partial}^{\Delta_{2}|1}(\partial_W)\le[
%(W\cdot \nabla_{Y_{\lambda}})^2 + \Delta_3 W^2\ri]G_{b \partial}^{\Delta_{3}|0}
\end{multline}
 From here we see that another natural relation arises between the OPE structures and interactions: 
\be\label{eq:i3}
V_{1,23}V_{2,13}: \quad  A_{1}^{\mu}A_{2}^{\nu}\partial_{(\mu}\partial_{\nu)}\phi_3 ~~\sim ~~A_{1}^{\mu}A_{2}^{\nu}\le(\nabla_{(\mu}\nabla_{\nu)}+ \Delta_3 g_{\mu\nu} \ri)\phi_3  ~.
 \ee
 where the sign $\sim$ here means that the relation is schematic: to rewrite interactions with partial derivatives as covariant derivatives, we are using homogeneity properties of fields in the embedding formalism in \eqref{eq:VVOstr}. In what follows we will keep most of our expressions in terms of partial derivatives.

Now let's consider building $G_{\Delta_1,\Delta_2,\Delta_3|1,1,0}$ starting from a geodesic diagram where we integrate over $\gamma_{13}$ instead of $\gamma_{12}$ (second diagram in Fig. \ref{fig:gwd3}).  The diagram with $\gamma_{12}$ already suggested as candidate interactions \eqref{eq:aap} and \eqref{eq:i3}. If we integrate those interactions over $\gamma_{13}$  we find\footnote{We are being schematic and brief in \eqref{eq:aa13}: it is implicit that we are using bulk--to--boundary propagators.}
\begin{align}\label{eq:aa13}
\int_{\gamma_{13}} A_{1}^{\mu}A_{2}^{\nu}\partial_{(\mu}\partial_{\nu)}\phi_3 = 0~,
\end{align}
and $A_{1}^{\mu}A_{2 \mu} \phi_3$  gives a linear combination of $V_{1,23}V_{2,13}$ and $H_{12}$. The identifications we made in \eqref{eq:aap} and \eqref{eq:i3} are obviously sensitive to the geodesic we select (there is a non-trivial kernel), and this is somewhat unsatisfactory.  We can partially overcome this pathology by considering a wider set of interactions. By inspection we find that the tensor structure $V_{1,23}V_{2,13}$ is simultaneously captured by  $\gamma_{13}$ and $\gamma_{12}$ by the interactions
\be\label{eq:vvaa}
V_{1,23}V_{2,13}: \qquad \alpha_1 A_1^{\nu}A_2^{\mu}  \partial_{\nu}\partial_{\mu} \phi_3 - \beta_1\le(
(\Delta_1+\Delta_2) \phi_3 \partial_{\mu} A^{\nu}_1 \partial_{\mu} A_1^{\nu} - 
(1+\Delta_1 \Delta_2) \phi_3 \partial_{\nu} A_1^{\mu} \partial_{\mu} A_2^{\nu} \ri) ~.
 \ee
The choice of geodesic affects the overall normalization, controlled by the choice of constants $\alpha_1$ and $\beta_1$. The terms multiplying $\beta_1$ when projected over $\gamma_{12}$, are proportional to the tensor structure $H_{12}$ and their coefficients are chosen such that they cancel each other. The interaction multiplying $\alpha_1$ is identically zero when integrated over $\gamma_{13}$.  To capture $H_{12}$ along both   $\gamma_{13}$ and $\gamma_{12}$  we just need
 \be \label{eq:hff}   
H_{12}:\qquad \phi_3 F_{1\mu\nu} F_2^{\mu\nu}~.
\ee
Here it is important  to note we are not using  $A_{1}^{\mu}A_{2 \mu} \phi_3$ as we did in \eqref{eq:aap}, and we still find the correct result when using $\gamma_{12}$. This is because there are many ways we can cast $H_{12}$ as bulk quantities along $\gamma_{12}$: the relation \eqref{eq:h123} is not unique. For instance, one can check that 
\begin{align}\label{eq:amb}
&G_{b\partial}^{\Delta_1|1}(Y_\lambda,P_1; \partial_W,Z_1) G_{b\partial}^{\Delta_2|1}(Y_\lambda,P_2;W,Z_2)\cr
&=-\frac{1}{2(\Delta_1+\Delta_2)}(\partial_W\cdot \partial_{Y'})(\partial_{W'}\cdot \partial_Y) G_{b\partial}^{\Delta_1|1}(Y',P_1;W',Z_1) G_{b\partial}^{\Delta_2|1}(Y,P_2;W,Z_2)\big|_{Y=Y'=Y_\lambda}\cr
&=-\frac{1}{2(1+\Delta_1\Delta_2)}(\partial_Y\cdot \partial_{Y'}) G_{b\partial}^{\Delta_1|1}(Y',P_1;\partial_{W},Z_1) G_{b\partial}^{\Delta_2|1}(Y,P_2;W,Z_2)\big|_{Y=Y'=Y_\lambda}
\end{align}
This type of relations are due to the projections over the geodesic, and they generate quite a bit of ambiguity as one tries to re-cast a given geodesic diagram as arising from a cubic interaction. Establishing relations such as  \eqref{eq:vvaa} and \eqref{eq:hff} are not fundamental, and their ambiguity is not merely due to integrating by parts or using equations of motion. In appendix \ref{app:tts} we provide some further examples on how to rewrite certain tensor structures as interactions, but we have not taken into account ambiguities such as those in \eqref{eq:amb}. Generalizing \eqref{eq:vvaa} and \eqref{eq:hff} for higher spin fields is somewhat cumbersome (but not impossible). We comment in the discussion what are the computational obstructions we encounter to carry this out explicit. 

\subsection{Basis of cubic interactions via Witten diagrams}\label{sec:st}

In the above we made use of our bulk differential operators to identify which interactions capture the suitable tensor structures that label the various correlation functions in the bulk. It is time now to compare with the results in \cite{Sleight2016b}. 

The most general cubic vertex among the symmetric-traceless fields of spin $J_i$ and mass $M_i$ ($i=1,2,3$) is a linear combination of interactions \cite{Joung2012,Taronna2012,Joung2013,Joung2013a}
\begin{align}\label{eq:v3}
V_{3}= \sum_{n_i=0}^{J_i}g(n_i) I_{J_1,J_2,J_3}^{n_1,n_2,n_3}(Y_i)|_{Y_i=Y}~,
\end{align}
 where $g(n_i)$ are arbitrary coupling constants, and
 \begin{multline}\label{eq:i1}
  I_{J_1,J_2,J_3}^{n_1,n_2,n_3}(Y_i) = {\cal Y}_1^{J_1-n_2-n_3}{\cal Y}_2^{J_2-n_3-n_1}{\cal Y}_3^{J_3-n_1-n_2}\\ \times {\cal H}_1^{n_1}{\cal H}_2^{n_2}{\cal H}_3^{n_3} \,{\cal T}_{J_1}(Y_1,W_1){\cal T}_{J_2}(Y_2,W_2){\cal T}_{J_3}(Y_3,W_3)~.
\end{multline}
Here ${\cal T}_{J_i}(Y_i,W_i)$ are polynomials in the embedding formalism that contain the components of the symmetric traceless tensor field in AdS. This cubic interaction is built out of six basic contractions which are defined as\footnote{As mentioned before all derivatives here are partial, but by using the homogeneity of  ${\cal T}_{J_i}(Y_i,W_i)$ one can relate them to covariant derivatives. }
\begin{align}\label{eq:cv3}
{\cal Y}_1 =\partial_{W_1}\cdot \partial_{Y_2}~,\quad {\cal Y}_2 =\partial_{W_2}\cdot \partial_{Y_3}~,\quad {\cal Y}_3 =\partial_{W_3}\cdot \partial_{Y_1}~,\cr
{\cal H}_1 =\partial_{W_2}\cdot \partial_{W_3}~,\quad {\cal H}_2 =\partial_{W_1}\cdot \partial_{W_3}~,\quad {\cal H}_3 =\partial_{W_1}\cdot \partial_{W_2}~.
\end{align}
For more details on the construction of this vertex we refer to \cite{Taronna2012}. What is important to highlight here are the following two features. First, $V_3$ is the most general interaction modulo field re-parametrization and total derivatives. Second, the number of terms in \eqref{eq:v3} is exactly the same as the number of independent structures in a CFT three point function \eqref{eq:n123}. 

The precise map between these interactions and tensor structures is in appendix A of \cite{Sleight2016b} (which is too lengthy to reproduce here). The first few terms give the following map:\footnote{Here the notation $\xrightarrow[{\rm bulk}]{}$ means that the identification between the interaction and tensor structure is done via a bulk integral, i.e. a three-point Witten diagram. Similarly,  $\xrightarrow[\gamma_{ij}]{}$ denotes an analogous integral over a geodesic.}
\begin{align}\label{eq:mapst}
  &I_{1,0,0}^{0,0,0} = A_{1}^\mu (\partial_\mu \phi_2)  \phi_3 \quad &\xrightarrow[{\rm bulk}]{}  &\quad V_{1,23}\cr 
  &I_{1,1,0}^{1,0,0} = A_{1}^\mu A_{2\mu}   \phi_3 \quad &\xrightarrow[{\rm bulk}]{}  &\quad \left((\Delta_1-\Delta_2)^2-\Delta_3^2\right) V_{1,23}V_{2,13} -(-2 \Delta_1 \Delta_2+\Delta_1+\Delta_2-\Delta_3)H_{12} \cr
 &I_{1,1,0}^{0,0,0} = A_{1}^\mu (\partial_\mu A_2^{\nu})  \partial_\nu \phi_3 \quad &\xrightarrow[{\rm bulk}]{}  &\quad (\Delta_1+\Delta_2-\Delta_3-2) V_{1,23}V_{2,13} +H_{12}
 \end{align}
In a nutshell this map is done by evaluating suitable Witten diagrams that capture three point functions and identify the resulting tensor structures. In appendix \ref{app:bwd} we derive specific examples to illustrate the mapping. Using this same basis of interactions and integrating them along $\gamma_{12}$ gives the following map
 \begin{align}\label{eq:mapstgeo}
  &I_{1,0,0}^{0,0,0} = A_{1}^\mu (\partial_\mu \phi_2)  \phi_3 \quad \xrightarrow[\gamma_{12}]{} \quad 0\cr 
  &I_{1,1,0}^{1,0,0} = A_{1}^\mu A_{2\mu}   \phi_3 \quad \xrightarrow[\gamma_{12}]{} \quad H_{12} \cr
 &I_{1,1,0}^{0,0,0} = A_{1}^\mu (\partial_\mu A_2^{\nu})  \partial_\nu \phi_3 \quad \xrightarrow[\gamma_{12}]{} \quad H_{12}
\end{align}
 Clearly there is a tension between the tensor structures we assign to an interaction if we use a regular Witten diagram versus a geodesic diagram. The mismatch is due to the fact that certain derivatives contracted along $\gamma_{ij}$ are null.  This reflects upon that a geodesic diagram is sensitive to the arrangement of derivatives which, for good reasons, are discarded in \eqref{eq:v3}.

 %The analogous map we found in section \ref{sec:vss}-\ref{sec:vvs} by studying the projection over  $\gamma_{12}$ (i.e. a geodesic that projects the bulk--to--boundary propagators of $A_1$ and $A_2$) was
%\begin{align}\label{eq:mapus}
 %&A_{1}^\mu (\partial_\mu \phi_2)  \phi_3 \quad \rightarrow \quad V_{1,23}\cr 
 %&A_{1}^\mu A_{2\mu}   \phi_3 \quad\rightarrow \quad H_{12} \cr
 %&A_{1}^\mu A_2^{\nu} \partial_\mu \partial_\nu \phi_3 \quad\rightarrow \quad  V_{1,23}V_{2,13} 
%\end{align}
%In \eqref{eq:mapst} and \eqref{eq:mapus} we are ignoring overall normalizations of the vertex and the map, since they are not crucial for what follows.
Some agreements do occur.  Let us reconsider the basis of interactions found by using geodesic interactions; from \eqref{eq:hff} we have (up to overall normalizations)
\begin{align}
\phi_3 F_{1\mu\nu} F_2^{\mu\nu}\quad \xrightarrow[\gamma_{ij}]{} \quad H_{12}
\end{align}
%\begin{align}
%&A_{1\nu}A_{2\mu}  \partial_{\nu}\partial_{\mu} \phi_3 
%-(\Delta_1+\Delta_2) \phi_3 \partial_{\mu} A_{1\nu} \partial_{\mu} A_{2\nu} + 
%(1+\Delta_1 \Delta_2) \phi_3 \partial_{\nu} A_{1\mu} \partial_{\mu} A_{2\nu} \quad \xrightarrow[\gamma_{ij}]{} \quad V_{1,23}V_{2,13}~,\cr
 %&~.
%\end{align}
If we use these interactions on Witten diagrams, we obtain exactly the same map
\begin{align}\label{eq:altmap}
%&A_{1\nu}A_{2\mu}  \partial_{\nu}\partial_{\mu} \phi_3 
%-(\Delta_1+\Delta_2) \phi_3 \partial_{\mu} A_{1\nu} \partial_{\mu} A_{2\nu} + 
%(1+\Delta_1 \Delta_2) \phi_3 \partial_{\nu} A_{1\mu} \partial_{\mu} A_{2\nu} \quad \xrightarrow[{\rm bulk}]{} \quad V_{1,23}V_{2,13}~,\cr
 %&
 \phi_3 F_{1\mu\nu} F_2^{\mu\nu}\quad \xrightarrow[{\rm bulk}]{} \quad H_{12}~.
\end{align}
The details of the computations leading to \eqref{eq:altmap} are shown in appendix \ref{app:bwd}. Moreover, we find that the interaction \eqref{eq:vvaa}, which is $V_{1,23}V_{2,13}$ for the geodesic Witten diagram, gives the same tensor structure if we integrate over the bulk, as shown in \eqref{eq:vvaabulk}. These relations indicate that it is possible to a have a compatible map among interactions in geodesic diagrams and Witten diagrams, even though there is disagreement at intermediate steps. However, from a bulk perspective the interaction selected in \eqref{eq:altmap} is not in any special footing relative to those in \eqref{eq:v3}.

\section{Conformal block decomposition of Witten diagrams}\label{sec:decom}

For a fixed cubic interaction, there is generically a mismatch among tensor structures captured by Witten diagrams versus geodesic Witten diagrams. In this section we will analyse how this affects the decomposition of four-point Witten diagrams in terms of geodesic diagrams. 

\begin{figure}
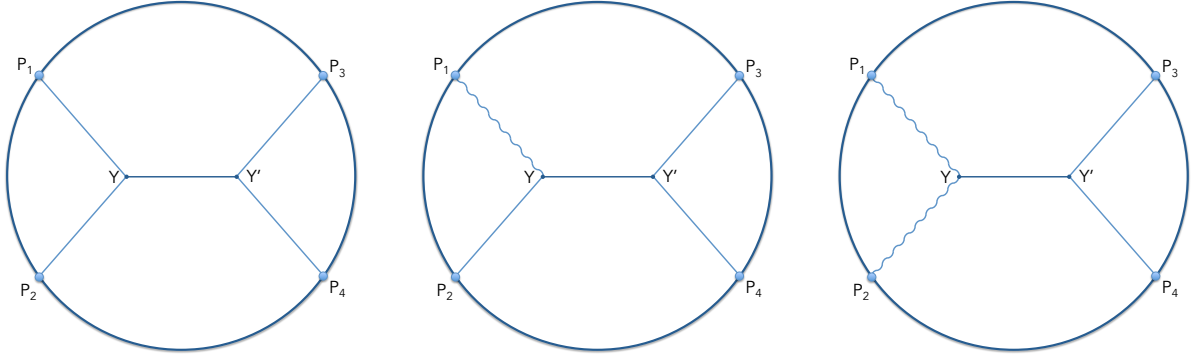

%\centering
{ \includegraphics[width=0.33\textwidth,page=12]{diagrams.pdf}
   \includegraphics[width=0.33\textwidth,page=13]{diagrams.pdf}
  \includegraphics[width=0.33\textwidth,page=14]{diagrams.pdf}
 }
  \caption{ Four-point exchange  Witten diagrams in AdS$_{d+1}$, where the exchanged field is a scalar field of dimension $\Delta$. The first diagram corresponds to ${\cal A}^{\rm Exch}_{{0,0,0,0}}$ in \eqref{eq:a4}, the second diagram to ${\cal A}^{\rm Exch}_{{1,0,0,0}}$ in \eqref{eq:a31}, and the third diagram to ${\cal A}^{\rm Exch}_{{1,1,0,0}}$ in \eqref{eq:a22}.}\label{fig:wd}
\end{figure}

Our discussion is based in the four-point exchange diagram for four scalars fields done in \cite{Hijano2016}, which we quickly review here. In Fig. \ref{fig:wd} we represent the exchange: all fields involved are scalars, where the external legs have dimension $\Delta_i$ and the exchange field has dimension $\Delta$. The corresponding Witten diagram is
\be\label{eq:a4}
{\cal A}^{\rm Exch}_{0,0,0,0}(P_i) = \int dY \int dY' G_{b\partial}^{\Delta_1|0}(Y, P_1)G_{b\partial}^{\Delta_2|0}(Y, P_2) G_{bb}^{\Delta|0}(Y, Y')G_{b\partial}^{\Delta_3|0}(Y', P_3)G_{b\partial}^{\Delta_4|0}(Y', P_4)~.
\ee
Here ``$dY$'' represents volume integrals in AdS$_{d+1}$. To write this expression as geodesic integrals, the crucial observation is that
\be\label{eq:gphi}
G_{b\partial}^{\Delta_1|0}(Y, P_1)G_{b\partial}^{\Delta_2|0}(Y, P_2) = \sum_{m=0}^\infty a_m^{\Delta_{1},\Delta_{2}} \varphi_m(\Delta_1,\Delta_2;Y)~,
\ee
where 
\bea\label{eq:phi12}
 \varphi_m(\Delta_1,\Delta_2;Y)\equiv \int_{\gamma_{12}} G_{b\partial}^{\Delta_1|0}(Y_\lambda, P_1)G_{b\partial}^{\Delta_2|0}(Y_\lambda, P_2)G_{bb}^{\Delta_m|0}(Y_\lambda, Y)~.
\eea
The field $ \varphi_m(Y)$ is a normalizable solution of  the Klein-Gordon equation with a source concentrated at $\gamma_{12}$ and mass  $M^2=\Delta_m(\Delta_m-d)$. The equality in \eqref{eq:gphi} holds provided one sets
\be
a_m^{\Delta_{1},\Delta_{2}} = {(-1)^m\over m!} {(\Delta_1)_m(\Delta_2)_m\over \beta_m (\Delta_1+\Delta_2+m-d/2)_m }~,\quad \Delta_m=\Delta_1+\Delta_2 +2m~.
\ee
 The constant $\beta_m$ soaks the choice of  normalizations used in \eqref{eq:phi12}.  Replacing \eqref{eq:phi12} twice in \eqref{eq:a4} gives
 \begin{multline}\label{eq:a41}
{\cal A}^{\rm Exch}_{{0,0,0,0}}(P_i) = \sum_{m,n} a_m^{\Delta_{1},\Delta_{2}}a_n^{\Delta_{3},\Delta_{4}}\int_{\gamma_{12}}\int_{\gamma_{34}}   G_{b\partial}^{\Delta_1|0}(Y_\lambda, P_1)G_{b\partial}^{\Delta_2|0}(Y_\lambda, P_2)G_{b\partial}^{\Delta_3|0}(Y'_{\lambda'}, P_3)G_{b\partial}^{\Delta_4|0}(Y'_{\lambda'}, P_4)\\
\times\int dY \int dY' G_{bb}^{\Delta_m|0}(Y_\lambda, Y)G_{bb}^{\Delta|0}(Y, Y') G_{bb}^{\Delta_n|0}(Y', Y'_{\lambda'})  ~.
 \end{multline}
 The integrals in the last line can be simplified by using
 \be
 G_{bb}^{\Delta|0}(Y, Y') = \langle Y | {1\over \nabla^2 - M^2} | Y'\rangle ~,\qquad \int dY | Y\rangle \langle Y | =1~,
 \ee
which leads to
\begin{multline}\label{eq:idg}
\int dY \int dY' G_{bb}^{\Delta_m|0}(Y_\lambda, Y)G_{bb}^{\Delta|0}(Y, Y') G_{bb}^{\Delta_n|0}(Y', Y'_{\lambda'}) =\\ {G_{bb}^{\Delta|0}(Y_{\lambda},Y'_{\lambda'})\over (M_\Delta^2-M^2_{m})(M_\Delta^2-M^2_{n})}+ {G_{bb}^{\Delta_m|0}(Y_{\lambda},Y'_{\lambda'})\over (M_m^2-M^2_{\Delta})(M_m^2-M^2_{n})} + {G_{bb}^{\Delta_n|0}(Y_{\lambda},Y'_{\lambda'})\over (M_n^2-M^2_{\Delta})(M_n^2-M^2_{m})}~.
\end{multline}
And hence the four-point exchange diagram for scalars is
\be
{\cal A}^{\rm Exch}_{{0,0,0,0}}(P_i) = C_{\Delta} {\cal W}_{\Delta | 0}(P_i) + \sum_{m} C_{\Delta_m} {\cal W}_{\Delta_m | 0}(P_i) +\sum_{n} C_{\Delta_n} {\cal W}_{\Delta_n | 0}(P_i)~,
\ee
where we organized the expression in terms of the geodesic integral that defines ${\cal W}_{\Delta | 0}$ in \eqref{eq:gwd1}; the coefficients $C_{\Delta}$ basically follow from the contributions in \eqref{eq:a41} and \eqref{eq:idg}.

\subsection{Four-point scalar exchange with one spin-1 field}

Now let's see how this decomposition will work when the external legs have spin. And the first non-trivial example is to just add a spin-1 field in one external leg and all other fields involved are scalar. The diagram is depicted in Fig. \ref{fig:wd}, and the integral expression is
\begin{multline}\label{eq:a31}
{\cal A}^{\rm Exch}_{1,0,0,0} = \int dY \int dY' G_{b\partial}^{\Delta_1|1}(Y, P_1,Z_1,\partial_W) \le(W\cdot \partial_Y G_{b\partial}^{\Delta_2|0}(Y, P_2)\ri) G_{bb}^{\Delta|0}(Y, Y')\\ \times G_{b\partial}^{\Delta_3|0}(Y', P_3)G_{b\partial}^{\Delta_4|0}(Y', P_4)~,
\end{multline}
where we used one of the vertex interactions in \eqref{eq:v3}. Using   \eqref{eq:gbds} and \eqref{eq:id345} we can rewrite this diagram in terms of the four-point scalar exchange \eqref{eq:a4} as
\be
{\cal A}^{\rm Exch}_{1,0,0,0}(\Delta_1, \Delta_2, \Delta_3,\Delta_4) = {2\Delta_2\over \Delta_1} \,D_{2\, 12} {\cal A}^{\rm Exch}_{0,0,0,0}(\Delta_1, \Delta_2+1, \Delta_3,\Delta_4)~,
\ee
and $D_{2\, 12}$ is defined in \eqref{eq:dij}. And from here the path is clear: using the geodesic decomposition and trading $D_{2 12}$ by $-{\cal D}_{2\, 12}(Y_\lambda)$ we obtain
\be
{\cal A}^{\rm Exch}_{1,0,0,0} =\tilde{C}_{\Delta} {\cal W}_{\Delta | 0}^{1,0,0,0} + \sum_{m} \tilde{C}_{\Delta_m} {\cal W}_{\Delta_m | 0}^{1,0,0,0} +\sum_{n} \tilde{C}_{\Delta_n} {\cal W}^{1,0,0,0}_{\Delta_n | 0}~,
\ee
with suitable constants $\tilde{C}$ and 
\begin{align}
{\cal W}_{\Delta | 0}^{1,0,0,0}(\Delta_1, \Delta_2, \Delta_3,\Delta_4)&= D_{2\, 12}{\cal W}_{\Delta | 0} (\Delta_1, \Delta_2+1, \Delta_3,\Delta_4)\cr
&= -\frac12 \int_{\gamma_{12}}\int_{\gamma_{34}}   G_{b\partial}^{\Delta_1|1}(Y_\lambda, P_1,Z_1,\partial_W)G_{b\partial}^{\Delta_2|0}(Y_\lambda, P_2)W\cdot \partial_{Y_\lambda} G_{bb}^{\Delta|0}(Y_\lambda,Y'_{\lambda'}) \cr &\qquad \qquad\times G_{b\partial}^{\Delta_3|0}(Y'_{\lambda'}, P_3)G_{b\partial}^{\Delta_4|0}(Y'_{\lambda'}, P_4)~,
\end{align}
 where we used \eqref{eq:voo}. It is interesting to note how the interaction gets slightly modified due to the cancellations that occur in the geodesic integrals: in \eqref{eq:a31} the derivative is acting on $G_{b\partial}^{\Delta_2|0}$, but the geodesic decomposition moves it to position of the exchanged field. 
  
 In this example it is also worth discussing the generalization of \eqref{eq:gphi}. Our decomposition of the bulk--to--boundary operators on position 1 and 2 reads
 \begin{align}
&G_{b\partial}^{\Delta_1|1}(Y, P_1,Z_1,\partial_W) W\cdot \partial_Y G_{b\partial}^{\Delta_2|0}(Y, P_2)\cr&\quad\qquad= 
{2\Delta_2\over \Delta_1} D_{2\,12}\le(G_{b\partial}^{\Delta_1|0}(Y, P_1)G_{b\partial}^{\Delta_2+1|0}(Y, P_2)\ri) \cr 
&\quad\qquad={2\Delta_2\over \Delta_1}  \sum_{m=0}^\infty a_m^{\Delta_1,\Delta_2+1} {\cal D}_{2\, 12}(Y) \varphi_m(\Delta_1,\Delta_2+1;Y)\cr
&\quad\qquad= -{\Delta_2\over \Delta_1} \sum_{m=0}^\infty a_m^{\Delta_{1},\Delta_{2}+1}  \int_{\gamma_{12}}  G_{b\partial}^{\Delta_1|1}(Y_\lambda, P_1,Z_1,\partial_W)G_{b\partial}^{\Delta_2|0}(Y_\lambda, P_2)\,W\cdot \partial_{Y_\lambda} G_{bb}^{\Delta_m|0}(Y_\lambda,Y)~.
\end{align}
 It is interesting to note the different interpretations one could give to the product $A^\mu_1\partial_\mu \phi_2$ (first line) in terms of resulting bulk fields.   Very crudely, from the third line one would like to say that  we just have a suitable differential operator acting on the field, while from the fourth line we would say that the product induces an interaction integrated along the geodesic. This type of decompositions of bulk fields would be interesting in the context of developing further a relation between an OPE expansion in the CFT to local bulk fields as done in \cite{Czech2016a,Boer2016,Guica2016}.
 
 \subsection{Four-point scalar exchange with two spin-1 fields}
 
 It is instructive as well to discuss an example with two spin-1 fields  as shown in the third diagram of Fig. \ref{fig:wd}. For sake of simplicity we will  use the cubic interaction $A_{1\mu} A^{\mu}_2 \phi$, which is part of the basis in \eqref{eq:v3}. The four-point exchange is 
 \begin{multline}\label{eq:a22}
{\cal A}^{\rm Exch}_{1,1,0,0} = \int dY \int dY' G_{b\partial}^{\Delta_1|1}(Y, P_1,Z_1,\partial_W) G_{b\partial}^{\Delta_2|1}(Y, P_2,Z_2,W) G_{bb}^{\Delta|0}(Y, Y')\\\times G_{b\partial}^{\Delta_3|0}(Y', P_3)G_{b\partial}^{\Delta_4|0}(Y', P_4)~.
\end{multline}
The new pieces are due to the presence of the spin-1 fields so we will focus on how to manipulate the propagators at position 1 and 2; the rest follows as in previous examples. Using \eqref{eq:gbds} allows us to remove the tensorial pieces in \eqref{eq:a22} and recast it in terms of tensor structures. For this case in particular we have
  %\begin{align}
%&G_{b\partial}^{\Delta_1|1}(Y, P_1,Z_1,\partial_W) G_{b\partial}^{\Delta_2|1}(Y, P_2,Z_2,W)\cr&~= 
%\le[\left((\Delta_1-\Delta_2)^2-\Delta_3^2\right) V_{1,23}V_{2,13} -(-2 \Delta_1 \Delta_2+\Delta_1+\Delta_2-\Delta_3)H_{12}\ri] G_{b\partial}^{\Delta_1|0}(Y, P_1)G_{b\partial}^{\Delta_2|0}(Y, P_2)\cr &~= 4 D_{1\, 12 }D_{1\, 21} G_{b\partial}^{\Delta_1+1|0}(Y, P_1)G_{b\partial}^{\Delta_2+1|0}(Y, P_2) -2 \Delta_1(1- \Delta_2)H_{12} G_{b\partial}^{\Delta_1|0}(Y, P_1)G_{b\partial}^{\Delta_2|0}(Y, P_2)~.
%&\qquad= 4\sum_{m=0}^\infty a_m (\Delta_1+1,\Delta_2+1){\cal D}_{1\, 12 }{\cal D}_{1\, 21}(Y) \varphi_m(\Delta_1+1,\Delta_2+1;Y)  \cr
%&\qquad\phantom{=}-2 \Delta_1(1- \Delta_2)\sum_{m=0}^\infty a_m (\Delta_1,\Delta_2)H_{12} \varphi_m(\Delta_1,\Delta_2;Y)\cr 
%&\qquad= \sum_{m=0}^\infty a_m  \int_{\gamma_{12}}  G_{b\partial}^{\Delta_1|1}(Y_\lambda, P_1,Z_1,\partial_W)G_{b\partial}^{\Delta_2|0}(Y_\lambda, P_2)\,W\cdot \partial_{Y_\lambda} G_{bb}^{\Delta_m|0}(Y_\lambda,Y)
%\end{align}
\begin{align}
G_{b\partial}^{\Delta_1|1}(Y, P_1,Z_1,\partial_W) G_{b\partial}^{\Delta_2|1}(Y, P_2,Z_2,W)&= {1\over \Delta_1 \Delta_2}\mathscr{D}_{P_1}(\partial_W,Z_1)\mathscr{D}_{P_2}(W,Z_2)G_{b\partial}^{\Delta_1|0}(Y, P_1) G_{b\partial}^{\Delta_2|0}(Y, P_2)\cr
&= {1\over \Delta_1 \Delta_2} \mathscr{D}_{P_1}(\partial_W,Z_1)\mathscr{D}_{P_2}(W,Z_2)\sum_{m=0}^\infty a_m^{\Delta_1,\Delta_2} \varphi_m(\Delta_1,\Delta_2;Y)~.\end{align}
From here we can relate the combination of $\mathscr{D}_P$'s acting on $\varphi_m$ to tensorial structures:
\begin{multline}\label{eq:g123}
\mathscr{D}_{P_1}(\partial_W,Z_1)\mathscr{D}_{P_2}(W,Z_2) \varphi_m(\Delta_1,\Delta_2;Y) =\\
-2 D_{1\, 12 }D_{1\, 21}  \int_{\gamma_{12}} G_{b\partial}^{\Delta_1+1|0}(Y_\lambda, P_1)G_{b\partial}^{\Delta_2+1|0}(Y_\lambda, P_2)G_{bb}^{\Delta_m|0}(Y_\lambda,Y)\\ - \Delta_1(1- \Delta_2)H_{12} \int_{\gamma_{12}} G_{b\partial}^{\Delta_1|0}(Y_\lambda, P_1)G_{b\partial}^{\Delta_2|0}(Y_\lambda, P_2)G_{bb}^{\Delta_m|0}(Y_\lambda,Y)~.
\end{multline}
This equality can be checked explicitly from the definitions of each term involved. A faster route is to infer it from the map given in \cite{Sleight2016a}: from \eqref{eq:mapst} we know the suitable structures in the interaction (which we just rewrote in terms of differential operators in \eqref{eq:g123}), and $\varphi_m$ behaves close enough to  a three point function that the map is unchanged. From here we can trade $D_{i\,jk}$ for ${\cal D}_{i\,jk}$, and then further use \eqref{eq:v1} and \eqref{eq:hint} to write them as smeared interactions. Without taking into account any normalizations, what we find for the contraction of two gauge fields decomposed in terms of geodesic integrals is
\begin{multline}\label{eq:g789}
G_{b\partial}^{\Delta_1|1}(Y, P_1,Z_1,\partial_W) G_{b\partial}^{\Delta_2|1}(Y, P_2,Z_2,W) \sim\\ \sum_m \int_{\gamma_{12}}  G_{b \partial}^{\Delta_{1}|1}(Y_\lambda;\partial_W)
 G_{b \partial}^{\Delta_{2}|1}(Y_\lambda;\partial_W)
(W\cdot \partial_{Y_{\lambda}})^2G_{bb}^{\Delta_{m}|0}(Y_\lambda,Y) \\+ \sum_m \int_{\gamma_{12}} G_{b\partial}^{\Delta_1|1}(Y_\lambda;\partial_{W})G_{b\partial}^{\Delta_2|1}(Y_\lambda;W) G_{bb}^{\Delta_m|0}(Y_\lambda,Y)~,
\end{multline}
where we are suppressing as well most of the variables in the propagators. This example illustrates how more interactions are needed when we decompose a Witten diagram in terms of geodesic diagrams; or in other words, how the product expansion of the bulk fields requires different interactions than those used in the direct evaluation of  a three point function. But more importantly, we should highlight that casting $G_{b\partial}^{\Delta_1|1}(Y, P_1,Z_1,\partial_W) G_{b\partial}^{\Delta_2|1}(Y, P_2,Z_2,W)$ as local interactions integrated along a geodesic is ambiguous. Consider as an example the last term in \eqref{eq:g789}. We could have written it in multiply ways due to the degeneracies shown in \eqref{eq:amb}: the product of two gauge fields could be casted as integrals of the interaction of $\phi A_\mu A^\mu$ or $\phi F_{\mu\nu} F^{\mu\nu}$ or similar contractions. And these interactions  are not related by equations of motion nor field redefinitions. As we discussed in section \ref{sec:st}, the identifications of gravitational interactions in a geodesic diagram is not unique and seems rather ad hoc. It  would be interesting to understand if there is a more fundamental principle underlying products such as those in \eqref{eq:g789}.

\subsection{Generalizations for scalar exchanges}\label{sec:gse}
In a nutshell, this is how we are decomposing a four-point scalar exchange Witten diagram in terms of geodesics diagrams: 
\begin{enumerate}
\item Consider a cubic interaction $I_{J_1,J_2,0}^{n_1,n_2,n_3}$ of the form \eqref{eq:i1}, where at position 1 and 2 we place bulk--to--boundary propagators and at position 3 we have a bulk--to--bulk propagator. From \eqref{eq:gbds} and \eqref{eq:id345} we will be able to strip off  the tensorial part of the interaction, i.e. schematically we will have
\be\label{eq:tr}
 I_{J_1,J_2,0}^{n_1,n_2,n_3} = \mathscr{D} \cdots\mathscr{D} \, I_{0,0,0}^{0,0,0}~.
\ee
Here ``$\mathscr{D} \cdots\mathscr{D}$'' symbolizes a chain of contractions of operators appearing in  \eqref{eq:gbds} and \eqref{eq:id345}, and the precise contraction depends on the interaction. The important feature is that $\mathscr{D} \cdots\mathscr{D}$ involves only derivatives with respect to $Z_i$ or $P_i$ (and not $Y$) which allows us to take this portion outside of the volume integral in a Witten diagram. Here $I_{0,0,0}^{0,0,0}$ is a cubic interaction for three scalars with the appropriate propagators used, i.e.
\be
I_{0,0,0}^{0,0,0}=G_{b\partial}^{\Delta_1|0}(Y, P_1)G_{b\partial}^{\Delta_2|0}(Y, P_2)G_{bb}^{\Delta|0}(Y,Y')~. 
\ee
%Note that we have chosen to set $J_3=0$ because we can very easily write the vertex as differential operators acting on a scalar precursor: all of the index structure can be placed on position 1 and 2 where we can manipulate the bulk--to--boundary propagators to cast it as \eqref{eq:tr}. We will discuss in the next section how to proceed when $J_3\neq0$, and potential obstacles in this case. 

\item The map among tensor structures and cubic interactions in \cite{Sleight2016a} implies that we will always be able to write the combination of $ \mathscr{D}$'s in terms of CFT operators:
\be
 \mathscr{D} \cdots\mathscr{D}  I_{0,0,0}^{0,0,0} = D \cdots D\, I_{0,0,0}^{0,0,0} ~.
\ee
This tells us which are the tensor structures appearing in the Witten diagram. 
\item Next we can rewrite $I_{0,0,0}^{0,0,0}$ as a sum over geodesic integrals via \eqref{eq:gphi}. This allows us to trade $D$ for our geodesic operators ${\cal D}(Y)$ as given in \eqref{eq:dopmap}:
\be
 \mathscr{D} \cdots\mathscr{D}  I_{0,0,0}^{0,0,0} = D \cdots D\, I_{0,0,0}^{0,0,0}= \mathcal{D} \cdots\mathcal{D}  I_{0,0,0}^{0,0,0} ~.
\ee
\item And if desired, we can as well write the action of $\mathcal{D}$ on $I_{0,0,0}^{0,0,0}$ as an interaction via the map in \eqref{eq:genint123}. This gives a more local description of the  OPE of the bulk fields in $ I_{J_1,J_2,0}^{n_1,n_2,n_3} $ in terms of smeared interactions along the geodesic. 

 \end{enumerate}
 
 A four-point exchange Witten diagram, where the exchange particle is a scalar field, is build out of two vertices of the form  $I_{J_1,J_2,0}^{n_1,n_2,n_3}$. So, keeping the loose schematic equalities, we can establish the following chain of equalities
 \begin{align}
 {\cal A}^{\rm exch}_{J_1,J_2,J_3,J_4} 
 &\sim \mathscr{D}_{\rm left} \mathscr{D}_{\rm right} {\cal A}^{\rm exch}_{0,0,0,0}\cr
 &\sim D_{\rm left} D_{\rm right} \, {\cal A}^{\rm exch}_{0,0,0,0}\cr
 & \sim \sum_m {\cal W}_{\Delta_m | 0}[\mathcal{D}_{\rm left}(Y_\lambda),\mathcal{D}_{\rm right}(Y'_{\lambda'})]~.
 \end{align}
where $ \mathscr{D}_{\rm left}$ corresponds to product of differential operators that recast the vertex to the left in terms boundary operators acting on position $(P_1,P_2)$, and the analogously for $\mathscr{D}_{\rm right}$ acting on $(P_3,P_4)$.

\subsection{Four-point spin exchanges} \label{sec:dswd}

\begin{figure}
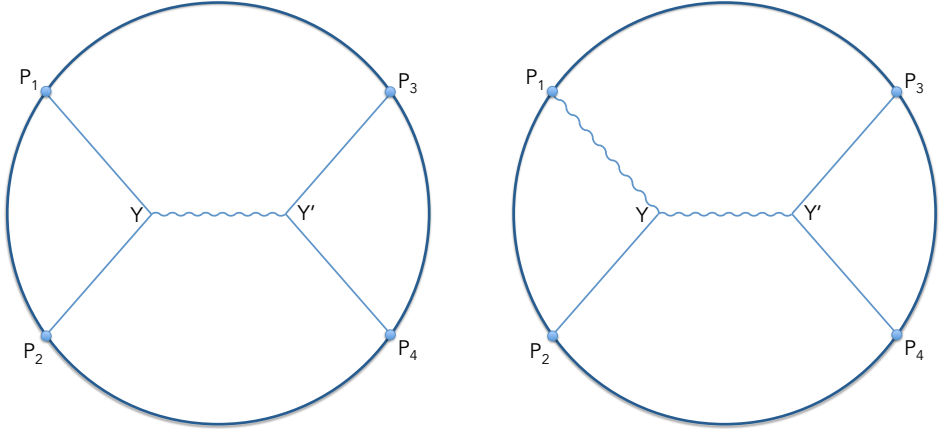

\centering
{ \includegraphics[width=0.4\textwidth,page=15]{diagrams.pdf}
   \includegraphics[width=0.4\textwidth,page=16]{diagrams.pdf}
 }
  \caption{ Four-point exchange  Witten diagrams in AdS$_{d+1}$, where the exchanged field is a symmetric tensor field of dimension $\Delta$ and spin $J$. In \eqref{eq:wds1}   and  \eqref{eq:wds11} we consider explicit examples where $J=1$ for the external and exchanged field.}\label{fig:wdspin}
\end{figure}

In this last portion we will  address examples where the exchanged field has spin, and illustrate how the  four-point exchange diagram can be decomposed in terms of the geodesic integrals. First consider the following Witten diagram
 \begin{multline}\label{eq:wds1}
{\cal A}^{\rm Exch | spin}_{0,0,0,0} = \int dY \int dY' G_{b\partial}^{\Delta_1|0}(Y, P_1) \partial_W \cdot\le( \partial_Y G_{b\partial}^{\Delta_2|0}(Y, P_2)\ri) G_{bb}^{\Delta|1}(Y, Y',W,\partial_{W'})\\\times W' \cdot \le(\partial_{Y'} G_{b\partial}^{\Delta_3|0}(Y', P_3)\ri)G_{b\partial}^{\Delta_4|0}(Y', P_4)~.
\end{multline}
In this diagram we are using the interaction $\phi_1 \partial_\mu \phi_2 A^\mu$ on both ends, and it is depicted in Fig. \ref{fig:wdspin}. The decomposition of \eqref{eq:wds1} in terms of geodesic integrals was done in \cite{Hijano2016} and we will not repeat it here. Next, let's consider a diagram where the field at position $P_2$ is a massive vector, i.e.
 \begin{multline}\label{eq:wds11}
{\cal A}^{\rm Exch | spin}_{0,1,0,0} = \int dY \int dY' G_{b\partial}^{\Delta_1|0}(Y, P_1) G^{\Delta_2|1}(Y,P_2;\partial_W,Z_2) G_{bb}^{\Delta|1}(Y, Y',W,\partial_{W'})\\\times W' \cdot \partial_{Y'} \le[G_{b\partial}^{\Delta_3|0}(Y', P_3)\ri]G_{b\partial}^{\Delta_4|0}(Y', P_4)~.
\end{multline}
This would be the second diagram in  Fig. \ref{fig:wdspin}, and we decided to use the interaction $\phi_1 A_{2}^{\mu} A_\mu$ for the cubic interaction on the left of the diagram. We can relate \eqref{eq:wds11} to \eqref{eq:wds1} by noticing the that the bulk--to--boundary operators satisfy the following series of identities
\begin{align}\label{eq:g567}
G_{b\partial}^{\Delta_1|0}(Y,P_1)G_{b\partial}^{\Delta_2|1}(Y,P_2;\partial_W,Z_2)&=
 { 1\over \Delta_2} \mathscr{D}_{P_2}(\partial_W, Z_2) G_{b\partial}^{\Delta_1|0}(Y,P_1)G_{b\partial}^{\Delta_2|0}(Y,P_2)\cr
&= {\Delta_2 - 1\over \Delta_2( \Delta_1 - 1)} D_{1\,21}\le[ {1\over P_{12}}  G_{b\partial}^{\Delta_1-1|0}(Y,P_1)(\partial_W\cdot \partial_Y)G_{b\partial}^{\Delta_2|0}(Y,P_2)\ri] \cr &
-{1\over \Delta_2-1} D_{2\,21}\le[ {1\over P_{12}}  G_{b\partial}^{\Delta_1|0}(Y,P_1) (\partial_W\cdot \partial_Y)G_{b\partial}^{\Delta_2-1|0}(Y,P_2)\ri]
 \end{align}
 Here we used \eqref{eq:gbds}, and then using the explicit polynomial dependence of $G_{b\partial}^{\Delta|0}(Y,P)$ to obtain the equality in the last line. It is interesting to note that we can now  write
 \begin{align}\label{eq:wds22}
 {\cal A}^{\rm Exch | spin}_{0,1,0,0} &= {\Delta_2 - 1\over \Delta_2( \Delta_1 - 1)} D_{1\,21}\le[ {1\over P_{12}} {\cal A}^{\rm Exch | spin}_{0,0,0,0} (\Delta_1-1, \Delta_2,\Delta_3,\Delta_4 )  \ri] \cr &
\quad-{1\over \Delta_2-1} D_{2\,21}\le[ {1\over P_{12}}  {\cal A}^{\rm Exch | spin}_{0,0,0,0} (\Delta_1, \Delta_2-1,\Delta_3,\Delta_4 ) \ri]
 \end{align}
 And from here we can proceed by  using the explicit decomposition of ${\cal A}^{\rm Exch | spin}_{0,0,0,0}$ in terms of geodesic diagrams in \cite{Hijano2016} and then trading $D_{i\,jk}$ by ${\cal D}_{i\,jk}$ (just as we we did  in the previous examples in this section).\footnote{Note that the factor of $P_{12}$ can be reabsorbed into bulk--to--boundary propagators projected along geodesics, i.e $${1\over P_{12}} G_{b\partial}^{\Delta_1|0}(Y_\lambda,P_1) G_{b\partial}^{\Delta_2|0}(Y_\lambda,P_2) = G_{b\partial}^{\Delta_1+1|0}(Y_\lambda,P_1)G_{b\partial}^{\Delta_2+1|0}(Y_\lambda,P_2)~.$$ Hence, as we cast \eqref{eq:wds22} as a sum over geodesic integrals, all terms will have a bulk interpretation.}
 
 The manipulations shown here are very explicit for the interaction we have selected, but they are robust and not specific to the example. We expect that in general we will be able to carry out a decomposition such as the one in \eqref{eq:g567} and have generalizations of \eqref{eq:wds22} without much difficulty.  It would be interesting to generalize this discussion and give a more systematic algorithm to decompose Witten diagrams in terms of geodesic integrals when the exchanged field has non-trivial spin.

\section{Discussion}\label{sec:disc}

%Interpretation and future directions
Our main result  was to give a systematic method to evaluate conformal partial waves  as geodesic integrals in AdS. From the CFT perspective, a spinning conformal partial wave is built from differential operators acting on the scalar conformal partial wave \cite{Costa2011}; here we presented the analog of these differential operators in AdS  and showed  that they reproduce the same effect as in the CFT. More succinctly, we established 
\be
W_{\Delta | l}^{l_1,l_2,l_3,l_4}(P_i;Z_i)= D_{\rm left}D_{\rm right} W_{\Delta | l}(P_1,P_2,P_3,P_4) = {\cal W}_{\Delta | l}[\mathcal{D}_{\rm left}(Y_\lambda),\mathcal{D}_{\rm right}(Y'_{\lambda'})]~,
\ee
where the last equality is a purely AdS object build out of geodesic integrals, while the left hand side are CFT quantities.  Our construction of $ {\cal D}$ and its equivalence to the CFT analog is given in section \ref{sec:maingwd}. We emphasise that this equivalence holds for any symmetric traceless field of spin $J_i$ and conformal dimension $\Delta_i$. We did not assume conservation of the fields, and the method works when all fields are of different species. 

The immediate  use of an object like ${\cal W}_{\Delta | l}$ is to evaluate correlation functions in holography. But relating the geodesic diagrams to regular Witten diagrams is a non-trivial task: interactions projected on geodesic integrals behave starkly different to interactions in  volume integrals as we have seen explicitly throughout section \ref{sec:interactions}. This mismatch between the two objects makes  more delicate the decomposition of a Witten diagram in terms of geodesics. We carry out explicit examples in section \ref{sec:decom}, and discuss the general relation when the exchanged field is a scalar. The strategy we adopt for this decomposition is inspired by the identities used in \cite{Sleight2016b}: one rewrites all tensorials properties of the interactions among bulk--to--boundary fields in terms of boundary operators acting on a scalar seed. This allows us to identify the CFT operators $D_{i\, jk}$, and  use then our bulk operators ${\cal D}_{i\, jk}$ to write a final answer in terms of a sum of geodesic integrals. As a result,  the set of cubic interactions needed to decompose a Witten diagram in terms of geodesic diagrams is larger than the basis in \eqref{eq:v3}.  Each individual geodesic integral is, however, much easier to evaluate.  

We have not discussed contact Witten diagrams here, but actually they can be treated very similarly as we did in section \ref{sec:decom}. The scalar case was done in \cite{Hijano2016}, so the task is to manipulate the vertex along the lines of the discussion in section \ref{sec:gse}:  the analog of \eqref{eq:tr} for a quartic interaction would allow us to identify the suitable tensor structures. Note that in a quartic interaction all propagators involved are bulk--to--boundary and hence we  can  strip off its tensorial features. We have not done this computation explicitly for quartic interactions, but a priori we do not expect major obstructions.   

In section \ref{sec:gwdspin} we gave a prescription on how to evaluate conformal partial waves via geodesic diagrams when the exchanged field has non-trivial spin. And the general strategy we have adopted in this work allowed us to relate the geodesic diagrams to Witten diagrams, as we discussed in section \ref{sec:dswd}.
 From this method it is not straightforward to infer the gravitational interaction, as we did in section \ref{sec:interactions}, with the main obstacle being the contractions of $dY^\mu/d\lambda$ appearing in the integrand. It might be interesting to improve our prescription, to make this connection more evident.  One reason it might be interesting to have this connection is to discuss conformal partial waves for anti-symmetric fields, and the differential operators that generate them. This is a case where the gravitational techniques can elucidate an organizational principle for those class of partial waves in the CFT. Until now the literature on conformal partial waves for non-symmetric structures is limited to \cite{Costa:2014rya,Echeverri:2015rwa,Rejon-Barrera:2015bpa,Iliesiu2016,Echeverri:2016dun,CostaHansenPenedonesEtAl2016,Costa2016,Schomerus:2016epl}, and finding a basis of differential operators that generates them would be very interesting. 

Another future direction that would be interesting to pursue is the addition of loops on the gravitational side. Very little is known about how to evaluate Witten diagrams beyond tree level, with the exception of the recent work in \cite{Aharony2016}. It would interesting to see how the geodesic diagram decomposition of a Witten diagram is  affected by the presence of loops: since the geodesic diagrams are conformal partial waves, we would expect that loops only modify the OPE coeffcients in the decomposition and the relation between masses in AdS and conformal dimensions in the CFT. Answering this question requires  understanding also how loops alter the geodesic diagram itself and its CFT interpretation. Since conformal partial waves are dictated purely by symmetries, we expect that its holographic dual is robust against loop corrections, and its relation to loop diagrams in AdS can be made clear and straightforward. We leave this line of questions for future work.

\section*{Acknowledgements}

It is a pleasure to thank Jan de Boer, Monica Guica, Dennis Karateev, and   Jamie Sully for useful discussions; and we are particularly grateful to Charlotte Sleight and Massimo Taronna for many helpful insights they gave us. This work is part of the Delta ITP consortium, a program of the NWO that is funded by the Dutch Ministry of Education, Culture and Science (OCW).
  A.C. and E.L. are supported by Nederlandse Organisatie voor Wetenschappelijk Onderzoek (NWO) via a Vidi grant.  F.R. is supported by the Mexican Consejo Nacional de Ciencia y Tecnolog\'{i}a.

\appendix
\section{More on CFT three point functions}\label{app:tensors}

Following the summary in section \ref{sec:cft}, in this appendix we give some more explicit information about the tensor structures appearing in the correlation functions.

In the main part of the text we have considered primaries operators with arbitrary  conformal dimensions. Unitarity CFTs have restrictions on the possible dimensions, and is it well known the unitarity bound is 
\be
\Delta \geq l+d-2~,\qquad  l \geq 1~.
\ee
The bound is saturated by conserved currents. The presence of a current implies as well further restrictions on the correlation functions, which can be implemented  in the index-free framework of \cite{Costa2011b}. Conservation of a symmetric traceless tensor requires that its divergence is zero; this implies that the entries of 
\be
R(P,Z)={1\over l(d/2+l-2) }(\partial \cdot D) T(Z,P) + O(Z^2,Z\cdot P)~,\qquad \partial \cdot D \equiv {\partial\over \partial P^A} D_A~,
\ee
are zero modulo pure gauge terms. Here the operator $D_A$ is the projector introduced in \eqref{eq:da1}. 

%\subsection*{Three point functions involving massive spin 2 fields.}

%Here we record a few examples of three point functions that involve a massive spin 2 field, and the constraints imposed by conservation. 
%\begin{align}
%\langle T(P_{1})\phi_{2}(P_{2})\phi_{3}(P_{3})\rangle
%= \lambda V_{1,23}^{2}T(\Delta_{1},\Delta_{2},\Delta_{3})
%\end{align}
%\begin{align}
%V_{1,23}^{2}T(\Delta_{1},\Delta_{2},\Delta_{3}) = \frac{4D_{1\, 12}^{2}T(\Delta_{1}+2,\Delta_{2},\Delta_{3})}{(\Delta_{12}-\Delta_{3})(2+\Delta_{12}-\Delta_{3})}
%\end{align}
%Conservation sets $\Delta_{1}=d$ and 
%\begin{align}
%\partial_{P_{1}}\cdot D \langle T(P_{1})\phi_{2}(P_{2})\phi_{3}(P_{3})\rangle=0
%\end{align}
%implies either $\lambda=0$ or $\Delta_{2}=\Delta_{3}$.
%\subsection*{TTS}
To see how conservation affects a three point function, consider the following two spin-2 fields and one scalar. This correlation function is the combination of three tensor structures:
\begin{align}
G_{\Delta_1,\Delta_2,\Delta_3|2,2,0}
= \left( \alpha H_{12}^{2} + \beta H_{12}V_{1,23}V_{2,31} + \gamma V_{1,23}^{2}V_{2,31}^{2} \right) 
	T(\Delta_{1},\Delta_{2},\Delta_{3})~,
\end{align}
where 
\begin{multline}
H_{12}V_{1,23}V_{2,31}T(\Delta_{1},\Delta_{2},\Delta_{3}) =\\ 
\frac{-4 H_{12}D_{1\,12}D_{1\,21}T(\Delta_{1}+1,\Delta_{2}+1,\Delta_{2})}{(\Delta_{12}-\Delta_{3})(\Delta_{12}+\Delta_{3})}
-
\frac{ H_{12}^{2}T(\Delta_{1},\Delta_{2},\Delta_{2})}{(\Delta_{12}-\Delta_{3})}
\end{multline}
\begin{multline}
V_{1,23}^{2}V_{2,31}^{2}T(\Delta_{1},\Delta_{2},\Delta_{3}) = 
\frac{16 H_{12}D_{1\,12}^{2}D_{1\,21}^{2}T(\Delta_{1}+2,\Delta_{2}+2,\Delta_{2})}{(\Delta_{12}-\Delta_{3})(\Delta_{12}+\Delta_{3})(2+\Delta_{12}+\Delta_{3})(-2+\Delta_{12}-\Delta_{3})}\\
+
\frac{16 H_{12}D_{1\,12}D_{1\,21}T(\Delta_{1}+1,\Delta_{2}+1,\Delta_{2})}{(\Delta_{12}-\Delta_{3})(\Delta_{12}+\Delta_{3})(-2+\Delta_{12}-\Delta_{3})}+\frac{2H_{12}^{2}T(\Delta_{1},\Delta_{2},\Delta_{2})}{(\Delta_{12}-\Delta_{3})(-2+\Delta_{12}-\Delta_{3})}
\end{multline}
Conservation implies $\Delta_{1}=\Delta_{2}=d$ and 
\begin{align}
\alpha = \frac{4h(h-1)(2h+1)-4\Delta_{3}h(2h-1)+\Delta_{3}^{2}(2h-1)}{2 \Delta_{3}(\Delta_{3}+2) (h-1)} \gamma~,
\end{align}
\begin{align}
\beta=-\frac{2+4h^{2}+\Delta_{3}-2h(\Delta_{3}+1)}{(h-1)(\Delta_{3}+2)} \gamma~,
\end{align}
where $h=d/2$. Further recent developments on properties of correlation functions for conserved currents can be found in \cite{Zhiboedov2012,Sleight2016a} and references within.

\section{Tensor structures in Witten diagrams }\label{app:bwd}

In this appendix we will evaluate three point Witten diagrams explicitly to illustrate how the tensor structures appear in the final answer.  We will focus on the following interactions: 
\be
 A_{1}^{\mu}\, \partial_{\mu}\partial_{\nu}\phi_2 \, A^{\nu}_{3}~,\quad  \partial_{\mu}A_{1}^{\nu}\,\phi_2\, \partial_{\mu}A^{\nu}_{3}~, \quad \partial_{\mu}A_{1}^{\nu}\,\phi_2\, \partial_{\nu}A^{\mu}_{3}~. 
 \ee
 We will do this by using the techniques in \cite{Sleight2016b,Sleight2016}, where they write the $J$ spinning bulk to boundary propagator and its derivatives in terms of the scalar propagators. This allows us to express the three point function of our interest in terms of scalar three point functions. In our case, we will just need the following identities for the spin-1 case, which follow from \eqref{eq:gbds} and \eqref{eq:id345}:
\begin{align}\label{eq:bdprop1}
 \Delta \,G_{b\partial}^{\Delta|1}(Y,P;W,Z)&= \mathscr{D}_{P}(W,Z) \, G_{b\partial}^{\Delta|0}(Y,P)~,\\
(W'\cdot \partial_Y) G_{b\partial}^{\Delta|1}(Y,P;W,Z)&=  \mathscr{D}'_{P}(W',W,Z) \, G_{b\partial}^{\Delta+1|0}(Y,P)~,\label{eq:bdprop2}
 \end{align} 
 where $\mathscr{D}_{P}$ are differential operators defined as
\begin{align}\label{eq:chop}
&\mathscr{D}_{P}(W,Z)= (Z \cdot W)\left( Z \cdot \frac{\partial}{\partial Z}- P \cdot \frac{\partial}{\partial P}\right)+(P\cdot W) \left(Z\cdot \frac{\partial}{\partial P}\right)\,, \\
& \mathscr{D}'_{P}(W',W,Z)= 2\left((Z \cdot W')(P \cdot W) +\Delta (P \cdot W')(Z\cdot W) +(P \cdot W')(P \cdot W)\left( Z\cdot \frac{\partial}{\partial P}\right)  \right)\,.\label{eq:chop2}
 \end{align} 
 These operators should not be confused with the $D_{1,2}$ CFT operators in \eqref{eq:dij} or with the bulk diferential operators ${\cal D}_{1,2}$ in \eqref{eq:dopsbulk}. 
 
 We start by evaluating a Witten diagram using the interaction  $A_{1}^{\mu}\, \partial_{\mu}\partial_{\nu}\phi_2 \, A^{\nu}_{3}$. We have
\be
\int dY  \, G_{b\partial}^{\Delta_1|1}(Y,P_1;\partial_{W_1},Z_1)  G_{b\partial}^{\Delta_3|1}(Y,P_3;\partial_{W_3},Z_3) (W_1\cdot \partial_Y)(W_3\cdot \partial_Y)\,G_{b\partial}^{\Delta_2|0}(Y,P_2) \,.
\ee
Here $dY$ denotes an integral over the volume of AdS. Using \eqref{eq:bdprop} and \eqref{eq:chop2} gives
\begin{align}
& \frac{4\Delta_{2}(\Delta_{2}+1)}{\Delta_{1}\Delta_{3}}\int dY  \, \mathscr{D}_{P_1}(\partial_{W_1},Z_1)G_{b\partial}^{\Delta_1|0}(Y,P_1) \mathscr{D}_{P_3}(\partial_{W_3},Z_3) G_{b\partial}^{\Delta_3|0}(Y,P_3) (W_1\cdot P_2)(W_3\cdot P_2)\,G_{b\partial}^{\Delta_2+2|0}(Y,P_2) \cr
&= \frac{4\Delta_{2}(\Delta_{2}+1)}{\Delta_{1}\Delta_{3}}\mathscr{D}_{P_1}(P_2,Z_1)  \mathscr{D}_{P_3}(P_2,Z_3) \int dY  \,G_{b\partial}^{\Delta_1|0}(Y,P_1) G_{b\partial}^{\Delta_3|0}(Y,P_3)\,G_{b\partial}^{\Delta_2+2|0}(Y,P_2)\cr
 &= \frac{4\Delta_{2}(\Delta_{2}+1)\mathsf{C}_{\Delta_{1},\Delta_{2}+2,\Delta_{3}} }{\Delta_{1}\Delta_{3}}\mathscr{D}_{P_1}(P_2,Z_1)   \mathscr{D}_{P_3}(P_2,Z_3) T(\Delta_1,\Delta_2+2,\Delta_3)\,,
\end{align}
where 
\begin{align}
\mathsf{C}_{\Delta_{1},\Delta_{2},\Delta_{3}} = g\frac{\pi^{h}}{2}\Gamma \left( \frac{\Delta_{1}+\Delta_{2}+\Delta_{3}-2h}{2} \right) \frac{\Gamma \left( \frac{\Delta_{1}+\Delta_{2}-\Delta_{3}}{2} \right)\Gamma \left(\frac{\Delta_{1}+\Delta_{3}-\Delta_{2}}{2} \right) \Gamma \left(\frac{\Delta_{2}+\Delta_{3}-\Delta_{1}}{2} \right)}{\Gamma(\Delta_{1})\Gamma(\Delta_{2})\Gamma(\Delta_{3})}\,.
\end{align}
 Notice that $\mathscr{D}_{P_1}(P_2,Z_1)=D_{2,12}$, and $\mathscr{D}_{P_3}(P_2,Z_3)=D_{2,32}$. Now, applying the differential operators to the scalar 3-point function we find that tensor structure corresponding to the previous diagram is the following linear combination:
\begin{multline}
A_{1}^{\mu}\, \partial_{\mu}\partial_{\nu}\phi_2 \, A^{\nu}_{3}\,:\\ \frac{\Delta_{2}(\Delta_{2}+1)(\Delta_{1}-\Delta_{2}+\Delta_{3}-2)\mathsf{C}_{\Delta_{1},\Delta_{2}+2,\Delta_{3}}\left(H_{13}+(\Delta_1-\Delta_2+\Delta_3 -2)\,\, V_{1,23} V_{3,21}\right)}{\Delta_{1}\Delta_{3}}\,.
\end{multline}

For the interaction  $\partial_{\mu}A_{1}^{\nu}\,\phi_2\, \partial_{\mu}A^{\nu}_{3}$ we have
\be
\int dY \,(\partial_{W'}\cdot \partial_Y) \, G_{b\partial}^{\Delta_1|1}(Y,P_1;\partial_{W},Z_1)({W'}\cdot \partial_Y) G_{b\partial}^{\Delta_3|1}(Y,P_3;{W},Z_3)\,G_{b\partial}^{\Delta_2|0}(Y,P_2) \,,
\ee
which using \eqref{eq:bdprop2} is equivalent to
\begin{align}
 &\mathscr{D}'_{P_1}(\partial_{W'},\partial_{W},Z_1) \mathscr{D}'_{P_3}(W',W,Z_3)\int dY \, G_{b\partial}^{\Delta_1+1|0}(Y,P_1)G_{b\partial}^{\Delta_3+1|0}(Y,P_3)\,G_{b\partial}^{\Delta_2|0}(Y,P_2)\cr
 &=\mathsf{C}_{\Delta_{1}+1,\Delta_{2},\Delta_{3}+1}\mathscr{D}'_{P_1}(\partial_{W'},\partial_{W},Z_1) \mathscr{D}'_{P_3}(W',W,Z_3)\,T(\Delta_1+1,\Delta_2,\Delta_3+1)\,.
 \end{align}
Contracting the $W$'s in the differential operators gives
 \begin{align}
 &\mathscr{D}'_{P_1}(\partial_{W'},\partial_{W},Z_1) \mathscr{D}'_{P_3}(W',W,Z_3)=(\Delta_1+\Delta_3)(Z_1\cdot P_3)(Z_3\cdot P_1)+\Delta_1\Delta_3 (Z_1\cdot Z_3)(P_1\cdot P_3)\cr
 &+(P_1\cdot P_3)\left( (Z_1\cdot \partial_{P_1})(Z_3\cdot \partial_{P_3})+(1+\Delta_3)(P_1\cdot Z_3) (Z_1\cdot \partial_{P_1})+(1+\Delta_1)(P_3\cdot Z_1)(Z_3\cdot \partial_{P_3})\right)\,,\nonumber
 \end{align}
which leads to the following identification
\begin{multline}
\partial_{\mu}A_{1}^{\nu}\,\phi_2\, \partial_{\mu}A^{\nu}_{3}\,:\\ \mathsf{C}_{\Delta_{1}+1,\Delta_{2},\Delta_{3}+1}\left( (\Delta_1-\Delta_2+\Delta_3-2\Delta_1\Delta_3) H_{13}-(\Delta_1-\Delta_2-\Delta_3)(\Delta_1+\Delta_2-\Delta_3)\,\, V_{1,23} V_{3,21}\right)\,.
\end{multline}
The interaction  $\partial_{\mu}A_{1}^{\nu}\,\phi_2\, \partial_{\nu}A^{\mu}_{3}$ is computed analogously as the previous, but with different $W$ contractions 
\begin{align}
 &\int dY \,(\partial_{W'}\cdot \partial_Y) \, G_{b\partial}^{\Delta_1|1}(Y,P_1;\partial_{W},Z_1)({W}\cdot \partial_Y) G_{b\partial}^{\Delta_3|1}(Y,P_3;{W'},Z_3)\,G_{b\partial}^{\Delta_2|0}(Y,P_2) \cr
 &=\mathsf{C}_{\Delta_{1}+1,\Delta_{2},\Delta_{3}+1}\mathscr{D}'_{P_1}(\partial_{W'},\partial_{W},Z_1) \mathscr{D}'_{P_3}(W,W',Z_3)\,T(\Delta_1+1,\Delta_2,\Delta_3+1)\,.
 \end{align}
This contraction of the differential operators gives
 \begin{align}
 \mathscr{D}'_{P_1}&(\partial_{W'},\partial_{W},Z_1) \mathscr{D}'_{P_3}(W',W,Z_3)=\Delta_1\Delta_3 (Z_1\cdot P_3)(Z_3\cdot P_1)+(\Delta_1+\Delta_3)(Z_1\cdot Z_3)(P_1\cdot P_3)\\
 &+(P_1\cdot P_3)\left( (Z_1\cdot \partial_{P_1})(Z_3\cdot \partial_{P_3})+(1+\Delta_3)(P_1\cdot Z_3) (Z_1\cdot \partial_{P_1})+(1+\Delta_1)(P_3\cdot Z_1)(Z_3\cdot \partial_{P_3})\right)\,,\nonumber
 \end{align}
which applying it to the scalar three point function gives
\begin{multline}
\partial_{\mu}A_{1}^{\nu}\,\phi_2\, \partial_{\nu}A^{\mu}_{3}\,:\\ \mathsf{C}_{\Delta_{1}+1,\Delta_{2},\Delta_{3}+1}\left( -(\Delta_1+\Delta_2+\Delta_3-2) H_{13}-(\Delta_1-\Delta_2-\Delta_3)(\Delta_1+\Delta_2-\Delta_3)\,\, V_{1,23} V_{3,21}\right)\,.
\end{multline}

Based on these three interactions, we can make the following map
\begin{align}\label{eq:vvaabulk}
&H_{13}:\qquad  \partial_{\mu}A_{1}^{\nu}\,\phi_2\, \partial_{\mu}A^{\nu}_{3}-\partial_{\mu}A_{1}^{\nu}\,\phi_2\, \partial_{\nu}A^{\mu}_{3} ~,\cr
&V_{1,23} V_{3,21}: \quad \alpha \,A_{1}^{\mu}\, \partial_{\mu}\partial_{\nu}\phi_2 \, A^{\nu}_{3}-(\Delta_1+\Delta_3)\partial_{\mu}A_{1}^{\nu}\,\phi_2\, \partial_{\mu}A^{\nu}_{3}+(1+\Delta_1\Delta_3)\partial_{\mu}A_{1}^{\nu}\,\phi_2\, \partial_{\nu}A^{\mu}_{3}~,
\end{align}
 where
 \begin{align}
 \alpha = \frac{(\Delta_{1}-1)(\Delta_{3}-1)(\Delta_{1}-\Delta_{2}+\Delta_{3})(2+\Delta_{1}-\Delta_{2}+\Delta_{3})}{(\Delta_{1}+\Delta_{2}-\Delta_{3})(\Delta_{1}-\Delta_{2}-\Delta_{3})}~.
 \end{align}
%\textcolor{red}{ to do: comparison with geodesic Witten diagrams, normalizations of propagators and 3-point function, rewrite this last part of the 'good interaccions' when this is done'.}
Modulo normalizations, this identification is compatible with the identification using geodesic diagrams \eqref{eq:vvaa} and \eqref{eq:hff}.

\section{ Tensor-tensor-scalar structures via geodesic diagrams}\label{app:tts}

Based on the two examples in sections \ref{sec:vss} and \ref{sec:vvs},  we can make a general identification between tensorial structures and a minimal set of gravitational interactions that will capture them for a fixed choice of the geodesic given by the first diagram in Fig. \ref{fig:gwd3}. We saw that the simplest way to identify $H_{12}$ in the bulk is by an interaction that contracts indices among symmetric tensors at position 1 and 2, and the $V$'s added derivatives on position 3 with suitable contractions on legs 1 and 2. Hence, it seems like  each tensor structure $H_{12}^{p}V_{1,23}^{q}V_{2,13}^{r}T(\Delta_{1},\Delta_{2},\Delta_{3})$ is reproduced by a geodesic integral of the form
\begin{align}\label{eq:TTOclaim}
\int_{\gamma_{12}}d\lambda \frac{\mathcal{H}_{1\lambda}(Z_{1},\partial_{W})^{q}\mathcal{H}_{1\lambda}(Z_{1},\partial_{W'})^{p}}{\Psi_{1\lambda}^{\Delta_{1}}}
\frac{\mathcal{H}_{2\lambda}(Z_{2},\partial_{W})^{r}\mathcal{H}_{2\lambda}(Z_{2},W')^{p}}{\Psi_{2\lambda}^{\Delta_{2}}}
(W\cdot \partial_{Y_{\lambda}})^{q+r}\Psi_{3\lambda}^{- \Delta_{3}}~.
\end{align}
This is a claim we can prove. The proof requires the following identities which are easily obtained by induction:
\begin{align}
(W\cdot \partial_{Y_{\lambda}})^{n}\Psi_{3\lambda}^{- \Delta_{3}}&=(-2)^{n}(-\Delta_{3}-n+1)_{n}(W\cdot P_{3})^{n}\Psi_{3\lambda}^{- \Delta_{3}-n}~,\cr
(\mathcal{H}_{i\lambda}(Z_{i},\partial_{W}))^{n}(W\cdot P_{3})^{l}&=(l-n+1)_{n}(W\cdot P_{3})^{l-n}\left(\sqrt{\frac{P_{i3}\Psi_{3\lambda}}{\Psi_{i\lambda}}}\mathcal{V}_{\partial\, i,3\lambda}(Z_{i})\right)^{n}~,\cr
 \mathcal{H}_{1\lambda}(Z_{1},\partial_{W'})^{p}\mathcal{H}_{2\lambda}(Z_{2},W')^{p}|_{\gamma_{12}}&= p! H_{12}^{p}~.
 \end{align} 
 Applying these to the integral gives 
 \begin{multline}
 2^{q+r}p! q!(-\Delta_{3}-q-r+1)_{q+r}(q+1)_{r}\left( \frac{P_{13}P_{23}}{P_{12}} \right)^{\frac{q+r}{2}}H_{12}^{p}V_{1,23}^{q}V_{2,13}^{r}\\
 \times\int_{\gamma_{12}}\Psi_{1\lambda}^{-\Delta_{1}}\Psi_{2\lambda}^{-\Delta_{2}}\Psi_{3\lambda}^{-\Delta_{3}-q-r}~,
 \end{multline}
 where we used 
  \begin{align}
 \sqrt{\frac{P_{i3}\Psi_{31}}{\Psi_{i1}}}\mathcal{V}_{\partial\, i,31}(Z_{i})=-\sqrt{\frac{P_{13}P_{23}}{P_{12}}}\begin{cases}
 	V_{1,23}\quad \mathrm{if}~i=1\\
 	V_{2,13}\quad \mathrm{if}~i=2
 \end{cases}
 \end{align}
 The remaining integral evaluates to 
 \begin{align}
 \int_{\gamma_{12}}\Psi_{1\lambda}^{-\Delta_{1}}\Psi_{2\lambda}^{-\Delta_{2}}\Psi_{3\lambda}^{-\Delta_{3}-q-r}=\frac{T(\Delta_{1},\Delta_{2},\Delta_{3}+q+r)}{c_{\Delta_{1}\Delta_{2}\Delta_{3}+q+r}}~,
 \end{align}
 by (\ref{eq:TPFdef}). Therefore (\ref{eq:TTOclaim}) results in 
  \be
 \frac{2^{q+r}p! q!(-\Delta_{3}-q-r+1)_{q+r}(q+1)_{r}}{c_{\Delta_{1}\Delta_{2}\Delta_{3}+q+r}}H_{12}^{p}V_{1,23}^{q}V_{2,13}^{r}T(\Delta_{1},\Delta_{2},\Delta_{3})~,
 \ee
 which completes the proof. Hence, from the analysis of the integrals over the geodesic $\gamma_{12}$ (which connects the fields with spin), we find the following identification
 \be\label{eq:genint123}
 H_{12}^{p}V_{1,23}^{q}V_{2,13}^{r}: \quad h_{1\mu_1\cdots\mu_p}^{ \phantom{1\mu_1\cdots\mu_p}\alpha_1\cdots\alpha_q } h_{2}^{~\mu_1\cdots\mu_p \beta_1\cdots\beta_r } \partial_{\alpha_1}\cdots \partial_{\alpha_q} \partial_{\beta_1}\cdots \partial_{\beta_r} \phi_3~.
 \ee
 As we have noticed in section \ref{sec:int} this identification is not unique. It is sensitive to the choice of geodesic, and moreover to redundancies that appear as derivatives are contracted along $\gamma_{12}$ (i.e. generalizations of \eqref{eq:amb}). 
%\section{Identities}
%\begin{align}
%\mathcal{D}_{1\, ij}\Psi_{k1}^{l}&=-l \sqrt{\frac{P_{ik}\Psi_{i1}}{\Psi_{k1}}}\Psi_{k1}^{l}\mathcal{V}_{\partial\, i,k1}\\
%\mathcal{D}_{2\, ij}\Psi_{k1}^{l}&=-l \sqrt{\frac{P_{jk}\Psi_{j1}}{P_{ij}}}\Psi_{k1}^{l-1}
%\left( \sqrt{P_{jk}\Psi_{i1}}\mathcal{V}_{\partial\, i,j1} -\sqrt{P_{ik}\Psi_{j1}}V_{i,jk} \right) 
%\end{align}
%\begin{multline}
%\mathcal{D}_{1\, ij}\mathcal{V}_{\partial\, k,l1}=\frac{1}{2\sqrt{\Psi_{k1}\Psi_{l1}}}
%\left( 
%-\sqrt{P_{kl}}\Psi_{i1} \left[ H_{ik}-2 \mathcal{V}_{\partial\, i,k1}\mathcal{V}_{\partial\, k,i1} \right] \right. \\
%\left.
%-\sqrt{\Psi_{i1}}
%\left[ \sqrt{P_{il}\Psi_{k1}}\mathcal{V}_{\partial\, i,l1} -\sqrt{P_{ik}\Psi_{l1}}\mathcal{V}_{\partial\, i,k1} \right] \mathcal{V}_{\partial\, k,l1}
 %\right) 
%\end{multline}

\bibliographystyle{JHEP}
\bibliography{GWD}

\end{document}